\documentclass[12pt]{article}

\usepackage[english]{babel}
\usepackage{amsmath}
\usepackage{amssymb}
\usepackage{amsthm,amsfonts}
\usepackage{dsfont}
\usepackage{tensor}
\usepackage[dvipsnames]{xcolor}
\usepackage[mathstyleoff]{breqn}
\usepackage{slashed}
\usepackage{stmaryrd}
\usepackage{enumerate}
\usepackage[normalem]{ulem}
\usepackage{tikz}
\usepackage[mathscr]{euscript}
\usepackage{mathrsfs}
\usepackage[numbers,sort&compress]{natbib}
\usepackage{hyperref}
\usepackage{amsthm}
\usepackage{svrsymbols}
\usepackage{phaistos}
\usepackage{mathtools}
\usepackage{stmaryrd} 
\usepackage{graphicx}

\newcommand{\bmat}{\left(\begin{array}}
\newcommand{\emat}{\end{array}\right)}

\def\gtrsim{\mathrel{\raise.3ex\hbox{$>$\kern-.75em\lower1ex\hbox{$\sim$}}}}

\def\-{\hphantom{-}}

\def\s2{\frac{1}{\sqrt2}}

\def\Dsl{\,\raise.15ex\hbox{/}\mkern-13.5mu D} 

\def\be{\begin{equation}}
\def\ee{\end{equation}}
\def\bea{\begin{eqnarray}}
\def\eea{\end{eqnarray}}

\newcommand{\nn}{\nonumber}

\begin{document}


\pagestyle{plain}

\makeatletter
\@addtoreset{equation}{section}
\makeatother
\renewcommand{\theequation}{\thesection.\arabic{equation}}
\pagestyle{empty}
\begin{center}
\ \

\vskip .5cm

\LARGE{\LARGE\bf Carrollian limit of NS-NS and Heterotic Supergravity\\[10mm]}
\vskip 0.1cm

\large{Romina Ballesteros$^{(a)}$, Eric Lescano$^{(b)}$ and Sergio Patiño-López$^{(c)}$ 
 \\[2mm]}

{\small
$^{(a)}$ Universidad San Sebastián, Facultad de Ingeniería,
\textit{Bellavista 7, Recoleta, Santiago, Chile.}\\[0.5mm]

$^{(b)}$ University of Wroclaw, Faculty of Physics and Astronomy,\\
\textit{Maksa Borna 9, 50-204 Wroclaw, Poland.}\\[0.5mm]

$^{(c)}$ Universidad Nacional Autónoma de México, Departamento de Física de Altas Energías,\\ Instituto de Ciencias Nucleares,
\textit{Apartado Postal 70-543, CdMx 04510, México.}
}

{\small \verb"romina.ballesteros@uss.cl, eric.lescano@uwr.edu.pl, fis.selp16@gmail.com"}\\[.3cm]

\small{\bf Abstract} \\[0.1cm]\end{center}

We construct the Carrollian limit of NS--NS and heterotic supergravity through an ultra-relativistic expansion of the fields. An appropriate scaling of the dilaton renders the measure finite and compensates the divergences arising from the NS-NS supergravity Lagrangian, giving a finite action as $w\rightarrow\infty$. We then extend the construction to heterotic supergravity (neglecting fermions) by incorporating the non-Abelian gauge field together with the Green--Schwarz (GS) mechanism. The resulting theory contains a finite gauge sector consistently coupled to gravity, and the GS mechanism for the Carrollian 1-form field can be trivialized imposing field redefinitions. Then, we investigate the Carrollian equations of motion by both expanding the relativistic equations and deriving them from a variational principle. We also show that the leading $\alpha'$-corrected $\hat{\rm Riem}^2$ contribution remains finite under a rescaling of the string parameter $\alpha'\rightarrow \frac{\alpha'_c}{w^2}$, opening further research towards the full four-derivative effective action. Finally, we discuss the potential connection with the worldsheet formalism of the Carrollian string theory.


\setcounter{page}{1}
\pagestyle{plain}
\renewcommand{\thefootnote}{\arabic{footnote}}
\setcounter{footnote}{0}

\tableofcontents

\section{Introduction}

Carrollian geometry has attracted considerable attention in recent years as one of the fundamental non-Lorentzian limits of relativistic physics. Originally introduced through the ultra-relativistic contraction of the Poincaré algebra \cite{Poincare1}-\cite{Poincare2}, Carrollian structures were for a long time regarded as mathematical curiosities associated with the formal limit in which the speed of light approaches zero. More recently, however, Carrollian geometry has emerged as a unifying framework in several areas of theoretical physics, including black-hole horizons \cite{BH1}-\cite{NearHorizon3}, flat-space holography \cite{FlatHolo1}-\cite{FlatHolo12}, celestial holography \cite{CeleHolo1}-\cite{CeleHolo5}, null reductions \cite{Null1}-\cite{Null4} and effective descriptions of ultra-relativistic gravitational systems \cite{Eff1}-\cite{Eff11}. These developments have established Carrollian geometry as a natural counterpart of Newton--Cartan geometry \cite{NR1}-\cite{NRST10}, describing two complementary non-Lorentzian regimes of relativistic theories \cite{New}.

Parallel to these developments, considerable progress has been achieved in understanding Carrollian gravity directly from relativistic gravitational theories. Rather than constructing Carrollian geometries by gauging kinematical symmetry algebras, one may instead perform systematic ultra-relativistic expansions of a relativistic action, leading to Carrollian gravitational theories with intrinsic affine connections, curvature tensors and dynamical field equations. Recent constructions have subsequently extended this idea to Carrollian supergravity models \cite{Sugra1}-\cite{Sugra6}. 

By contrast, the ultra-relativistic limit of string effective actions remains comparatively unexplored. The low-energy dynamics of closed strings is governed by the NS--NS sector of supergravity, whose bosonic field content consists of the spacetime metric, the Kalb--Ramond 2-form and the dilaton. Unlike pure gravity, these additional fields play an essential role in string theory and significantly enrich the underlying geometric structure. Understanding how these fields behave under Carrollian contractions is therefore a necessary step towards the construction of Carrollian effective theories describing the ultra-relativistic regime of string theory. Such a construction is also expected to provide a geometric framework for investigating Carrollian strings, tensionless limits and other non-Lorentzian sectors of string theory from the spacetime point of view.

The situation becomes even richer in heterotic string theory, where the NS--NS sector couples to non-Abelian Yang--Mills gauge fields through the Green--Schwarz mechanism \cite{GS}. The resulting supergravity contains both gauge and antisymmetric tensor fields (we neglect the fermionic sector in this work), whose gauge transformations are intrinsically intertwined through the Chern--Simons modification of the Kalb--Ramond field strength. It is therefore far from obvious whether this structure survives the Carrollian limit, or whether the Green--Schwarz mechanism admits a consistent non-Lorentzian realization. Understanding this question is essential for extending Carrollian effective theories beyond the universal NS--NS sector and towards more realistic string compactifications.

The purpose of this work is to investigate the Carrollian limit of the bosonic NS--NS and heterotic supergravity actions. Starting from the relativistic theories, we introduce systematic large-$w$ expansions for the metric, Kalb--Ramond, dilaton and Yang--Mills fields, identifying the scaling of each field required for the action to possess a finite ultra-relativistic limit. The resulting theory naturally defines a Carrollian geometry whose fundamental variables consist of a clock 1-form, a degenerate spatial metric, a Carrollian 1-form inherited from the Kalb--Ramond field and a spatial 2-form. The corresponding affine connection arises directly as the finite contribution of the Levi--Civita connection. This leads to intrinsic non-metricities that are uniquely determined by compatibility with the parent relativistic geometry, similarly to the recent formulation of non-relativistic bosonic supergravity of \cite{EL}.

For the NS--NS sector we derive the complete finite Carrollian action together with its diffeomorphism and gauge symmetries. We show in section \ref{NSNS}  that the resulting theory admits a fully covariant formulation in terms of Carrollian curvature tensors and gauge-invariant field strengths, providing a dynamical realization of a string-inspired Carrollian geometry. We then extend the construction to heterotic supergravity by considering the ultra-relativistic limit of the Yang--Mills sector and of the Chern--Simons terms in section \ref{Heterotic}. In particular, we show that the Carrollian limit naturally separates the gauge transformations of the 1-form inherited from the Kalb--Ramond field from those of the spatial 2-form, allowing the introduction of a gauge-invariant Carrollian vector while preserving the non-trivial Green--Schwarz transformation of the 2-form. Interestingly, the Green--Schwarz transformation for the Carrollian 1-form can be trivialized, as we explicitly show. We derive the complete finite heterotic action and discuss its gauge-covariant formulation.

In section \ref{EOM} we compute the equations of motion of the theory. One option is to expand the relativistic equations, which are provided in equations (\ref{EOMinicial})-(\ref{EOMfinal}) and impose geometric constraints which provide a consistent truncation. The other option is to directly compute the explicit variations from the finite Lagrangian. The latter requires the inclusion of Lagrange multipliers to be able to satisfy the constitutive relations of the Carrollian geometry. We provide the variations for the Carrollian dilaton gravity case in the appendix \ref{App}. 

Finally in section \ref{Discussion} we include a detailed comparison with the non-relativistic limit and an explicit check of the $\hat{\rm Riem}^2$ contributions towards the inclusion of the $\alpha'$-corrections. In particular we prove that this contribution is finite under the rescaling $\alpha'\rightarrow \frac{\alpha'_c}{w^2}$. This indicates that in the case $\hat H_{\mu \nu \rho}=0$, the bosonic supergravity is finite up to four-derivative terms. The same applies for the bosonic sector of the heterotic supergravity, when $\bar H_{\mu \nu \rho}=0$.
In the last part of this section we also discuss potential relations to the worldsheet formalism. We believe that the framework developed here provides a first step towards a systematic formulation of Carrollian effective theories for string theory. Besides extending recent developments in Carrollian gravity beyond the purely gravitational sector, our results suggest that non-Lorentzian limits of supergravity possess a considerably richer geometric structure than their relativistic counterparts, opening new directions for the study of Carrollian string backgrounds, non-Lorentzian dualities and generalized Carrollian geometries.

\section{NS-NS Carrollian gravity}
\label{NSNS}
We start by imposing the following Carrollian ansatz for the metric (and inverse metric) tensor, \footnote{In this work we develop all our constructions in metric formalism, avoiding the decomposition of the Lorentz group. However, we provide a brief discussion of the boost symmetry in Section \ref{Boost}.}
\bea
\hat g_{\mu \nu} & = & h_{\mu \nu} - \frac{1}{w^2} \tau_{\mu} \tau_{\nu} \, , \\
\hat g^{\mu \nu} & = & h^{\mu \nu} - w^2 \tau^{\mu} \tau^{\nu} \, , 
\label{Metric}
\eea
where $w=\frac{1}{c}$. The Carrollian constitutive relations are
\bea
\label{cr1}
\tau_{\mu} h^{\mu \nu} & = & \tau^{\mu} h_{\mu \nu} = 0 \, , \\
\tau_{\mu} \tau^{\mu} & = & 1 \, , \\
\tau_{\mu} \tau^{\rho} + h_{\mu \nu} h^{\nu \rho} & = & \delta_{\mu}^{\rho} \, .
\label{cr3}
\eea
The Carrollian ansatz for the relativistic B-field contains two fundamental fields, a 2-form $b_{\mu \nu}$ and a 1-form $A_{\mu}$, i.e.,  
\bea
\hat B_{\mu \nu} = b_{\mu \nu} + 2 \tau_{[\mu} A_{\nu]} \, , 
\label{B}
\eea
where $A_{\mu} \tau^{\mu}=0$. The ansatz for the dilaton is given by
\bea
\hat \phi = \alpha \ln{w} + \varphi \, ,
\label{scaling}
\eea
with $\alpha$ determined by  
\bea
\sqrt{-\hat g} e^{-2 \hat \phi} =  \Omega_{c} w^{-2\alpha-1} e^{-2\varphi} = \Omega_{c} w^{-2}  e^{-2\varphi} \, ,  
\eea
and $\Omega_c$ the finite part of the square root of the determinant of the metric. Then $\alpha=\frac12$. Particularly, throughout the text we will use the combination $\int d^{10}x \Omega_c e^{-2\varphi} \hat L^{(2)}$ to represent the finite (supergravity) actions, where the supraindex in the Lagrangian means the order of the contribution in powers of $w$.

The fundamental geometric variables are the Carrollian clock 1-form $\tau_\mu$ and the degenerate spatial metric $h_{\mu\nu}$, together with the dilaton $\varphi$, the Carrollian 1-form $A_\mu$ and the spatial 2-form $b_{\mu\nu}$ inherited from the Kalb--Ramond field, these fields completely characterize the finite theory obtained after taking the ultra-relativistic limit. It is a straightforward computation to prove that the previous expansion leads to a finite Carrollian action upon considering $w\rightarrow\infty$,
\bea
S_{\rm NS-NS} = \int d^{10}x \sqrt{-\hat g} e^{-2 \hat \phi} (\hat R + 4 \partial_{\mu} \hat \phi \partial^{\mu} \hat \phi - \frac{1}{12} {\hat H}^2)
\eea
where $\hat H_{\mu \nu \rho} = 3 \partial_{[\mu} \hat B_{\nu \rho]}$. Since the measure $\sqrt{-\hat g} e^{-2 \hat \phi}$ produces $\frac{1}{w^2}$ contributions, the finite action is given by the $w^2$ contributions of $\hat R + 4 \partial_{\mu} \hat \phi \partial^{\mu} \hat \phi - \frac{1}{12} {\hat H}^2$.  

\subsection{Symmetry transformations and covariant derivatives}
At this point, it is straightforward to compute the finite Carrollian supergravity action. However, we are interested in write it in covariant form. For this reason, in this section we construct an affine connection and covariant derivatives with respect to infinitesimal diffeomorphisms.

\subsubsection{Symmetry rules}
The symmetry transformations of the fundamental fields are given by
\bea
\delta_{\xi} \tau_{\mu} & = &  \xi^{\rho} \partial_{\rho}\tau_{\mu} +  \partial_{\mu}{\xi^{\rho}} \tau_{\rho} \, , \\
\delta_{\xi} \tau^{\mu} & = & \xi^{\rho} \partial_{\rho} \tau^{\mu} - \partial_{\rho}\xi^{\mu} \tau^{\rho} \, , 
\\
\delta_{\xi} h_{\mu \nu} & = & \xi^{\rho} \partial_{\rho} h_{\mu \nu} + 2 \partial_{(\mu|} \xi^{\rho} h_{|\nu) \rho} \, , \\
\delta_{\xi} h^{\mu \nu} & = & \xi^{\rho} \partial_{\rho}{h^{\mu \nu}} – 2 \partial_{\rho}{\xi^{(\mu}} h^{ \nu) \rho}  \, , \\
\delta_{\xi} \varphi & = & \xi^{\rho} \partial_{\rho} \varphi \, , 
\\
\delta_{\xi} A_{\mu} & = & \xi^{\rho} \partial_{\rho}A_{\mu} +  \partial_{\mu}{\xi^{\rho}} A_{\rho} \, , \\
\delta_{\xi,\zeta} b_{\mu \nu} & = & \xi^{\rho} \partial_{\rho} b_{\mu \nu} + 2 \partial_{[\mu|} \xi^{\rho} b_{\rho|\nu]} + 2 \partial_{[\mu} \zeta_{\nu]} \, ,
\eea
where $\xi^{\mu}$ and $\zeta_{\mu}$ are generic parameters under infinitesimal diffeomorphism and b-shift transformations, respectively.  All fields transform covariantly under diffeomorphisms, and this is the symmetry that will allow us to rewrite the finite action in a fully covariant form. In addition, we include a b-shift symmetry for the Carrollian 2-form.

\subsubsection{Connections and non-metricities}
 We define the following torsionless affine connection under diffeomorphisms,
\bea
\Gamma_{\mu \nu}^{\rho} & = & \frac12 \tau^{\rho \sigma} \Big(2 \partial_{(\mu}(\tau_{\nu)} \tau_{\sigma}) - \partial_{\sigma}(\tau_{\mu} \tau_{\nu}) \Big) + \frac12 h^{\rho \sigma} \Big(2 \partial_{(\mu} h_{\nu) \sigma}  - \partial_{\sigma}(h_{\mu \nu}) \Big) \, .
\eea
Similarly to what happens in the non-relativistic limit \cite{EL}, the $w^2$ and $w^{-2}$ terms in the connections transforms as tensors, so one can define the Carrollian afine connection just by taking the ${\cal O}(1)$ contribution from
\bea
\hat \Gamma_{\mu \nu}^{\rho} = w^2 \hat \Gamma^{(2) \rho}_{\mu \nu} + \Gamma_{\mu \nu}^{\rho} + w^{-2} \hat \Gamma^{(-2)\rho}_{\mu \nu}  \, .
\eea
The covariant derivatives are therefore defined as \footnote{In order to move the formulation to the vielbein formalism, one should shift this connection and introduce the realizations of the spin connection, keeping the compatibility with the relativistic theory.}
\bea
\nabla_{\mu} \tau_{\nu} & = & \partial_{\mu} \tau_{\nu } - \Gamma_{\mu \nu}^{\rho} \tau_{\rho} \, , \\
\nabla_{\mu} \tau^{\nu}& = & \partial_{\mu} \tau^{\nu} + \Gamma_{\mu \sigma}^{\nu} \tau^{\sigma} \, , \\
\nabla_{\mu} h_{\nu \rho} & = & \partial_{\mu} h_{\nu \rho} - 2 \Gamma_{\mu (\nu}^{\sigma} h_{\rho) \sigma} \, , \\
\nabla_{\mu} h^{\nu \rho} & = & \partial_{\mu} h^{\nu \rho} + 2 \Gamma_{\mu \sigma}^{(\nu} h^{\rho) \sigma} \, . 
\eea
The Riemann tensor can be easily computed from the commutator of the covariant derivatives acting on an arbitrary vector $v^{\mu}$,
\bea
[\nabla_{\mu},\nabla_{\nu}] v^{\rho} = R^{\rho}{}_{\epsilon \mu \nu} v^{\epsilon} \, , 
\eea
giving the usual expression
\bea
R^{\rho}{}_{\epsilon \mu \nu}(\tau,h) = 2 \partial_{[\mu} \Gamma_{\nu] \epsilon}^{\rho} + 2 \Gamma_{[\mu| \alpha}^{\rho} \Gamma_{|\nu] \epsilon}^{\alpha} \, .
\eea
The previous expression should not be confused with the relativistic Riemann tensor, \footnote{Our convention is to use the hat symbol on the relativistic quantities, and the only exception will be $\bar H_{\mu \nu \rho}$ for the curvature of the relativistic Kalb-Ramond field in the heterotic setup.}
\bea
\hat R^{\rho}{}_{\epsilon \mu \nu} = 2 \partial_{[\mu} \hat \Gamma_{\nu] \epsilon}^{\rho} + 2 \hat \Gamma_{[\mu| \alpha}^{\rho} \hat \Gamma_{|\nu] \epsilon}^{\alpha} \, , 
\eea
which can be further decomposed as
\bea
\hat R^{\rho}{}_{\epsilon \mu \nu} = w^{4} \hat R^{(4)\rho}{}_{\epsilon \mu \nu} + w^{2} \hat R^{(2)\rho}{}_{\epsilon \mu \nu} + \hat R^{(0)\rho}{}_{\epsilon \mu \nu} + w^{-2} \hat R^{(-2)\rho}{}_{\epsilon \mu \nu} + w^{-4} \hat R^{(-4)\rho}{}_{\epsilon \mu \nu} \, .
\label{Riemannfull}
\eea
We demand compatibility with $\hat g_{\mu \nu}$ and $\hat g^{\mu \nu}$ to fix the value of the intrinsic non-metricities of this formulation,
\bea
\nabla_{\mu} (\tau_{\nu} \tau_{\rho}) & = & Q^{(\tau,\tau)}_{\mu \nu \rho} \, , \\
\nabla_{\mu} (\tau^{\nu} \tau^{\rho}) & = & Q^{(\tau,\tau)}_{\mu}{}^{\nu \rho} \, ,  \\
\nabla_{\mu} h_{\nu \rho} & = & Q^{(h)}_{\mu \nu \rho} \, , \\
\nabla_{\mu} h^{\nu \rho} & = & Q^{(h^{-1})}_{\mu}{}^{\nu \rho} \, . 
\eea
The previous quantities are given by:
\bea
\label{metricity1}
Q^{(\tau,\tau)}_{\mu \nu \rho} & = & h^{\epsilon \sigma} (\tau_{\mu} \nabla_{(\nu} \tau_{\sigma} + \tau_{(\nu} \nabla_{\mu} \tau_{\sigma} - \tau_{\mu} \nabla_{\sigma} \tau_{(\nu} - \tau_{(\nu} \nabla_{\sigma} \tau_{\mu}) h_{\epsilon \rho)} \, , \\
Q^{(\tau,\tau)}_{\mu}{}^{\nu \rho} & = & - \tau^{\sigma} \tau^{(\nu}(\nabla_{\mu} h_{\epsilon \sigma} + \nabla_{\epsilon} h_{\mu \sigma} - \nabla_{\sigma}h_{\mu \epsilon})h^{\rho) \epsilon} \, ,  \\
Q^{(h)}_{\mu \nu \rho} & = & \tau^{\sigma}(\nabla_{\mu}  h_{(\nu| \sigma} + \nabla_{(\nu|} h_{\mu \sigma} - \nabla_{\sigma} h_{\mu (\nu})\tau_{\rho)}\\
Q^{(h^{-1})}_{\mu}{}^{\nu \rho} & = & - 2 h^{(\nu|\sigma} (\tau_{(\mu} \nabla_{\epsilon)} \tau_{\sigma} - \tau_{(\mu} \nabla_{\sigma)} \tau_{\epsilon)} \tau^{\rho)} \tau^{\epsilon}\, , 
\label{metricity4}
\eea
and they are fixed to preserve the relativistic metric compatibility $\hat \nabla_{\mu} \hat g_{\nu \rho} = \hat \nabla_{\mu} \hat g^{\nu \rho}=0$.

\subsection{Covariant NS-NS action under the Carrollian limit}
Let us start by exploring the contribution given by $\hat R^{(2)}$ in (\ref{Riemannfull}). The covariant form of this contribution is
\bea
\hat R^{(2)} = \hat R^{(2)}_{\epsilon \nu} h^{\epsilon \nu} - \tau^{\epsilon} \tau^{\nu} \hat R^{(0)}_{\epsilon \nu}
\eea
where
\bea
\hat R^{(2)}_{\epsilon \nu} = - \nabla_{[\mu} \Big[\tau^{\mu} \tau^{\sigma} (\nabla_{\epsilon} h_{\nu] \sigma} + \nabla_{\nu]} h_{\epsilon \sigma} - \nabla_{\sigma} h_{\nu] \epsilon}) \Big]
\eea
and
\bea
\hat R^{(0)}_{\epsilon \nu} = && R_{\epsilon \nu} + \frac12 \tau^{\mu} \tau^{\sigma} h^{\alpha \beta} (\nabla_{[\mu} h_{\alpha \sigma} + \nabla_{\alpha} h_{[\mu \sigma} - \nabla_{\sigma} h_{[\mu \alpha}) (\tau_{\epsilon} \nabla_{\nu]} \tau_{\beta} - \tau_{\epsilon} \nabla_{\beta} \tau_{\nu]} + \tau_{\nu]} \nabla_{\epsilon} \tau_{\beta} - \tau_{\nu]} \nabla_{\beta} \tau_{\epsilon}) \nn \\ &&
+ \frac12 h^{\mu \sigma} \tau^{\alpha} \tau^{\beta} (\tau_{[\mu} \nabla_{\alpha} \tau_{\sigma} - \tau_{[\mu} \nabla_{\sigma} \tau_{\alpha} + \tau_{\alpha} \nabla_{[\mu} \tau_{\sigma} - \tau_{\alpha} \nabla_{\sigma} \tau_{[\mu}) (\nabla_{[\epsilon} h_{\nu] \beta} + \nabla_{\nu]} h_{\epsilon \beta} - \nabla_{\beta} h_{\nu] \epsilon}) \, . \nn \\
\eea
From the previous expression we conclude that the Ricci contribution to the Lagrangian contains a $R_{\epsilon \nu} \tau^{\epsilon} \tau^{\nu}$ term, plus other terms containing covariant derivatives of the fundamental fields. 

The $w^2$-terms for the dilaton is $- \frac{1}{4} \partial_{\mu}\varphi \partial_{\nu}\varphi \tau^{\mu} \tau^{\nu}$ while the $w^2$-terms for the $\hat H$ contribution are 
\bea
-\frac{1}{12}[{\hat H}^2]^{(2)} & = &  \frac12 A_{\mu} A_{\nu} \nabla_{\rho}{\tau_{\sigma}} \nabla_{\gamma}{\tau_{\epsilon}} \tau^{\sigma} \tau^{\epsilon} h^{\mu \nu} h^{\rho \gamma} + \frac12 A_{\mu} A_{\nu} \nabla_{\rho}{\tau_{\sigma}} \nabla_{\gamma}{\tau_{\epsilon}} \tau^{\rho} \tau^{\gamma} h^{\mu \nu} h^{\sigma \epsilon} \nn \\ && + A_{\mu} \nabla_{\nu}{A_{\rho}} \nabla_{\sigma}{\tau_{\gamma}} \tau^{\gamma} h^{\mu \rho} h^{\nu \sigma} - \frac12 A_{\mu} A_{\nu} \nabla_{\rho}{\tau_{\sigma}} \nabla_{\gamma}{\tau_{\epsilon}} \tau^{\rho} \tau^{\gamma} h^{\mu \sigma} h^{\nu \epsilon} \nn \\ && + A_{\mu} A_{\nu} \nabla_{\rho}{\tau_{\sigma}} \nabla_{\gamma}{\tau_{\epsilon}} \tau^{\rho} \tau^{\epsilon} h^{\mu \sigma} h^{\nu \gamma} - A_{\mu} A_{\nu} \nabla_{\rho}{\tau_{\sigma}} \nabla_{\gamma}{\tau_{\epsilon}} \tau^{\rho} \tau^{\epsilon} h^{\mu \nu} h^{\sigma \gamma} \nn \\ && - A_{\mu} \nabla_{\nu}{A_{\rho}} \nabla_{\sigma}{\tau_{\gamma}} \tau^{\sigma} h^{\mu \rho} h^{\nu \gamma} + A_{\mu} \nabla_{\nu}{A_{\rho}} \nabla_{\sigma}{\tau_{\gamma}} \tau^{\sigma} h^{\mu \nu} h^{\rho \gamma} \nn \\ && - \frac12 A_{\mu} A_{\nu} \nabla_{\rho}{\tau_{\sigma}} \nabla_{\gamma}{\tau_{\epsilon}} \tau^{\sigma} \tau^{\epsilon} h^{\mu \rho} h^{\nu \gamma} - A_{\mu} \nabla_{\nu}{A_{\rho}} \nabla_{\sigma}{\tau_{\gamma}} \tau^{\gamma} h^{\mu \nu} h^{\rho \sigma} \nn \\ && + A_{\mu} \nabla_{\nu}{\tau_{\rho}} \tau^{\rho} \tau^{\sigma} h^{\mu \gamma} h^{\nu \epsilon} h_{\sigma \gamma \epsilon} - A_{\mu} \nabla_{\nu}{\tau_{\rho}} \tau^{\nu} \tau^{\sigma} h^{\mu \gamma} h^{\rho \epsilon} h_{\sigma \gamma \epsilon} \nn \\ && + \frac12 \nabla_{\mu}{A_{\nu}} \nabla_{\rho}{A_{\sigma}} h^{\mu \rho} h^{\nu \sigma} - \frac12 \nabla_{\mu}{A_{\nu}} \nabla_{\rho}{A_{\sigma}} h^{\mu \sigma} h^{\nu \rho} \nn \\ && - \nabla_{\mu}{A_{\nu}} \tau^{\rho} h^{\mu \sigma} h^{\nu \gamma} h_{\rho \sigma \gamma} + \frac14 \tau^{\mu} \tau^{\nu} h^{\rho \sigma} h^{\gamma \epsilon} h_{\mu \rho \gamma} h_{\nu \sigma \epsilon} \, ,
\eea
where we have defined the gauge invariant 3-form as, 
\bea
h_{\mu \nu \rho}=3 \partial_{[\mu} b_{\nu \rho]} \, .
\eea

Therefore, the full Carrollian NS-NS gravity is given by
\bea
S_{\rm NS-NS} = \int d^{10}x \Omega_{c} e^{-2 \varphi} (\hat R^{(2)} - 4 \partial_{\mu} \varphi \partial_{\nu} \varphi \tau^{\mu} \tau^{\nu} - \frac{1}{12} [\hat {H}^2]^{(2)}) \, .
\label{actionNSNS}
\eea

\subsection{The Carrollian geometry underlying the NS--NS sector}

The previous construction provides a complete Carrollian limit of the bosonic NS--NS sector of supergravity, including the metric, Kalb--Ramond field and dilaton. Before extending the analysis to heterotic supergravity, it is worthwhile to summarize the geometric structure emerging from the resulting theory. 

A distinctive feature of the present construction is that the affine connection is not postulated independently. Instead, it is obtained directly as the finite contribution in the large-$w$ expansion of the Levi--Civita connection. Consequently, the associated non-metricities are uniquely determined by requiring compatibility with the parent relativistic metric before taking the Carrollian limit. The resulting connection therefore inherits its covariant structure directly from the relativistic theory, providing a geometric framework in which the curvature tensors and covariant derivatives appearing in the action are naturally defined. 

This construction complements recent developments on generalized Carrollian limits, which have mainly focused on kinematical symmetries, particle dynamics and null reductions. Here, instead, the Carrollian geometry is obtained directly from the relativistic NS--NS supergravity action, leading to a fully interacting and covariant effective theory. Having established this geometric framework, we now proceed to extend the construction to heterotic supergravity, where the Yang--Mills fields and the Green--Schwarz mechanism introduce additional geometric structures beyond those present in the pure NS--NS sector. While apriori there is no reason to believe that all these contributions will survive the $w\rightarrow\infty$ limit, we will show in the affirmative in the next section.

\section{Heterotic Carrollian gravity}
\label{Heterotic}
In this section we extend our results to the heterotic case, neglecting the fermionic sector. For this reason, we will keep the metric-like formulation of the previous section. The present construction can be naturally extended to the vielbein formulation, including the spin connections arising from the decomposition of the Lorentz group in order to compatibilize the fermionic degrees of freedom. 

The expansion for the gauge field $\hat A_{\mu}^{i}$ also splits in two fundamental fields, $a_{\mu}{}^{i}$ and $\chi^{i}$, i.e.,
\bea
\hat A_{\mu}^{i} = a_{\mu}{}^{i} + \tau_{\mu} \chi^{i}
\label{A}
\eea
where $a_{\mu}{}^{i}$ is a gauge connection which satisfies $a_{\mu}{}^{i} \tau^{\mu}=0$ 
and $\chi^i$ is a gauge tensorial shift.

The heterotic supergravity action is now given by
\bea
S_{\rm het} = \int \sqrt{-\hat g} e^{-2 \hat \phi} (\hat R + 4 \partial_{\mu} \hat \phi \partial^{\mu} \hat \phi - \frac{1}{12} {\bar H}^2 - \frac14 \hat F_{\mu \nu i} \hat F^{\mu \nu i}) \, ,
\eea
where the curvature of the non-Abelian gauge field is defined as,\footnote{We observe that $f^i{}_{jk}$ are the gauge structure constants of the relativistic theory.}
\bea
\hat F_{\mu \nu}{}^{i} = 2 \partial_{[\mu} \hat A_{\nu]}{}^{i} - f^{i}{}_{jk} \hat A_{\mu}{}^{j} \hat A_{\nu}{}^{k} \, ,
\eea
and the curvature for the Kalb-Ramond field now contains Chern-Simons contributions,
\be
\bar H_{\mu\nu\rho}=3\left(\partial_{[\mu}\hat B_{\nu\rho]}- \hat C_{\mu\nu\rho}^{(g)}\right)\, , \label{barH}
\ee 
with the Chern-Simons 3-form defined by
\be
\hat C_{\mu\nu\rho}^{(g)}= \hat A^i_{[\mu}\partial_\nu \hat A_{\rho]i}-\frac13 f_{ijk} \hat A_\mu^i \hat A_\nu^j \hat A_\rho^k \, .
\ee
The inclusion of the Chern-Simons term in the definition of the curvature of the Kalb-Ramond field ensures that the new curvature is gauge invariant. The extra terms compensate the Green-Schwarz transformation of the Kalb-Ramond field,
\bea
\delta_{ \lambda} \hat B_{\mu\nu} & = & -(\partial_{[\mu} \lambda^i) \hat A_{\nu] i} \, ,
\eea
where $\lambda^{i}$ is a generic paramter of gauge transformation of the heterotic gauge group. 
\subsection{Gauging the symmetries and the covariant derivative}
\subsubsection{Gauge symmetry rules and Green-Schwarz mechanism}

We find the following gauge rules when taking the limit $w\rightarrow \infty$, 
\bea
\delta_{\lambda} a_{\mu}{}^{i} & = & \partial_\mu \lambda^i + f^i{}_{jk} \lambda^j a_{\mu}{}^{k}\, , \label{gaugetransa} \\
\delta_{\lambda} \chi^i & = & f^i{}_{jk} \lambda^j \chi^k\, , \label{gaugetransalpha} \\ 
\delta_{ \lambda} b_{\mu\nu} & = & -(\partial_{[\mu} \lambda^i) a_{\nu] i} \, , \\ 
\delta_{ \lambda} A_{\mu} & = & \frac12 (\partial_{[\mu} \lambda^i) \chi_{i}  \label{gaugetransA} \, , 
\eea
while the remaining fields do not transform under gauge transformations. We observe that $a_\mu{}^i$ transforms as the gauge connection of the Carrollian geometry, $\chi^i$ transforms as a gauge vector, $b_{\mu\nu}$ transforms in a non-covariant way, emulating a Green-Schwarz mechanism and $A_{\mu}$ also transforms in a Green-Schwarz fashion. While in the relativistic case these transformations are non-ambiguous and cannot be removed by field redefinitions, when one considers the Carrollian limit it is possible to redefine the field $A_{\mu}$ to trivialize its Green-Schwarz mechanism, 
\bea
\bar A_{\mu} = A_{\mu} - \frac12 a_{\mu}{}^{i} \chi_i \, .
\label{barA}
\eea
The new $\bar A_{\mu}$ field is gauge invariant, i.e. $\delta_{\lambda} \bar A_{\mu}=0$. One could be tempted to do the same with $b_{\mu \nu}$, as it happens in the non-relativistic heterotic formulation of \cite{Eheterotic}, but in this case the transformation for the $b_{\mu \nu}$ field is unambiguous and cannot be trivialized as in the relativistic case.

\subsubsection{Gauging the covariant derivative}

Considering an arbitrary gauge vector $v^{i}$, upon taking the Carrollian limit, the partial derivative acting on this vector is not covariant. Therefore we need to define a covariant derivative for the gauge symmetry in the following way,
\bea
\nabla_{\mu} v^{i} = \partial_{\mu} v^{i} - f^{i}{}_{jk} a_{\mu}{}^{j} v^{k} \, .
\eea 
The curvature for the gauge connection $a_\mu{}^i$ follows the same form as the relativistic case, i.e., 
\bea
f_{\mu \nu}{}^{i} = 2 \partial_{[\mu} a_{\nu]}{}^{i} - f^{i}{}_{jk} a_{\mu}{}^{j} a_{\nu}{}^{k} \, .
\eea
The field $\chi^i$ is already covariant, so we do not need to construct its curvature. For the b-field we also construct its curvature by mimiking the relativistic case,
\bea
\label{newcurva}
\bar h_{\mu \nu \rho} =  h_{\mu \nu \rho} - 3 c^{(g)}_{\mu \nu \rho}.
\eea
where $c^{(g)}_{\mu \nu \rho} = a_{[\mu}{}^{i} \partial_\nu a_{\rho]}{}^{i}-\frac13 f_{ijk} a_\mu{}^i a_\nu{}^j a_\rho{}^k \, $.
\subsection{Covariant heterotic action under the Carrollian limit}
It is a long but straightforward computation to prove that the bosonic sector of the heterotic supergravity action is finite after consider $w\rightarrow\infty$. The $w^2$ contributions to the Yang-Mills term are given by,
\bea
-\frac14[F_{\mu \nu i} F^{\mu \nu i}]^{(2)} & = & \frac12 \tau^{\mu} \tau^{\nu} f_{\mu \rho}{}^{i} f_{\nu \sigma i} h^{\rho \sigma}   + \chi_{i} \nabla_{\mu}{\tau_{\nu}} \tau^{\mu} \tau^{\rho} f_{\rho \sigma}{}^{i} h^{\nu \sigma} - \nabla_{\mu}{\chi^{i}} \tau^{\nu} f_{\nu \rho i} h^{\mu \rho} \nn \\ && - \chi^{i} \nabla_{\mu}{\tau_{\nu}} \tau^{\nu} \tau^{\rho} f_{\rho \sigma i} h^{\mu \sigma} + \frac12 \nabla_{\mu}{\chi_{i}} \nabla_{\nu}{\chi^{i}} h^{\mu \nu}  + \chi^{i} \nabla_{\mu}{\chi_{i}} \nabla_{\nu}{\tau_{\rho}} \tau^{\rho} h^{\mu \nu} \nn \\ && + \frac12 \chi^{2} \nabla_{\mu}{\tau_{\nu}} \nabla_{\rho}{\tau_{\sigma}} \tau^{\nu} \tau^{\sigma} h^{\mu \rho} + \frac12 \chi^{2} \nabla_{\mu}{\tau_{\nu}} \nabla_{\rho}{\tau_{\sigma}} \tau^{\mu} \tau^{\rho} h^{\nu \sigma} - \chi_{i} \nabla_{\mu}{\chi^{i}} \nabla_{\nu}{\tau_{\rho}} \tau^{\nu} h^{\mu \rho} \nn \\ && - \chi^{2} \nabla_{\mu}{\tau_{\nu}} \nabla_{\rho}{\tau_{\sigma}} \tau^{\mu} \tau^{\sigma} h^{\nu \rho}  \, ,
\eea
while the ${\bar H}^2$ contribution is
\bea
&& -\frac{1}{12}[{\bar H}]^{(2)}  =  \frac12 \bar{A} _{\mu} \bar{A} _{\nu} \nabla_{\rho}{\tau_{\sigma}} \nabla_{\gamma}{\tau_{\epsilon}} \tau^{\sigma} \tau^{\epsilon}  h^{\mu \nu} h^{\rho \gamma} + \frac12 \bar{A} _{\mu} \bar{A} _{\nu} \nabla_{\rho}{\tau_{\sigma}} \nabla_{\gamma}{\tau_{\epsilon}} \tau^{\rho} \tau^{\gamma}  h^{\mu \nu} h^{\sigma \epsilon} 
\nn \\ && 
+ \bar{A} _{\mu} \nabla_{\nu}{\bar{A} _{\rho}} \nabla_{\sigma}{\tau_{\gamma}} \tau^{\gamma}  h^{\mu \rho} h^{\nu \sigma} - \frac12 \bar{A} _{\mu} \bar{A} _{\nu} \nabla_{\rho}{\tau_{\sigma}} \nabla_{\gamma}{\tau_{\epsilon}} \tau^{\rho} \tau^{\gamma}  h^{\mu \sigma} h^{\nu \epsilon} 
+ \bar{A} _{\mu} \bar{A} _{\nu} \nabla_{\rho}{\tau_{\sigma}} \nabla_{\gamma}{\tau_{\epsilon}} \tau^{\rho} \tau^{\epsilon}  h^{\mu \sigma} h^{\nu \gamma} 
\nn \\ && 
- \bar{A} _{\mu} \bar{A} _{\nu} \nabla_{\rho}{\tau_{\sigma}} \nabla_{\gamma}{\tau_{\epsilon}} \tau^{\rho} \tau^{\epsilon}  h^{\mu \nu} h^{\sigma \gamma} \nn 
- \bar{A} _{\mu} \nabla_{\nu}{\bar{A} _{\rho}} \nabla_{\sigma}{\tau_{\gamma}} \tau^{\sigma}  h^{\mu \rho} h^{\nu \gamma} 
+ \bar{A} _{\mu} \nabla_{\nu}{\bar{A} _{\rho}} \nabla_{\sigma}{\tau_{\gamma}} \tau^{\sigma}  h^{\mu \nu} h^{\rho \gamma} 
\nn \\ && 
- \frac12 \bar{A} _{\mu} \bar{A} _{\nu} \nabla_{\rho}{\tau_{\sigma}} \nabla_{\gamma}{\tau_{\epsilon}} \tau^{\sigma} \tau^{\epsilon}  h^{\mu \rho} h^{\nu \gamma} - \bar{A} _{\mu} \nabla_{\nu}{\bar{A} _{\rho}} \nabla_{\sigma}{\tau_{\gamma}} \tau^{\gamma}  h^{\mu \nu} h^{\rho \sigma} 
- \frac12 \bar{A} _{\mu} \chi^{2} \nabla_{\nu}{\tau_{\rho}} \nabla_{\sigma}{\tau_{\gamma}} \tau^{\nu}  h^{\mu \gamma} h^{\rho \sigma} 
\nn \\ && 
+ \frac12 \bar{A} _{\mu} \chi_{j} \nabla_{\nu}{\tau_{\rho}} \tau^{\nu}  f_{\sigma \gamma}{}^{j} h^{\mu \sigma} h^{\rho \gamma}  
- \frac12 \bar{A} _{\mu} \chi^{2} \nabla_{\nu}{\tau_{\rho}} \nabla_{\sigma}{\tau_{\gamma}} \tau^{\rho}  h^{\mu \sigma} h^{\nu \gamma} + \frac12 \bar{A} _{\mu} \chi^{2} \nabla_{\nu}{\tau_{\rho}} \nabla_{\sigma}{\tau_{\gamma}} \tau^{\nu}  h^{\mu \sigma} h^{\rho \gamma} 
\nn \\ && 
+ \frac12 \bar{A} _{\mu} \chi^{2} \nabla_{\nu}{\tau_{\rho}} \nabla_{\sigma}{\tau_{\gamma}} \tau^{\rho}  h^{\mu \gamma} h^{\nu \sigma} - \frac12 \bar{A} _{\mu} \chi_{j} \nabla_{\nu}{\tau_{\rho}} \tau^{\rho}  f_{\sigma \gamma}{}^{j} h^{\mu \sigma} h^{\nu \gamma} 
+ \bar{A} _{\mu} \nabla_{\nu}{\tau_{\rho}} \tau^{\rho} \tau^{\sigma}  h^{\mu \gamma} h^{\nu \epsilon} \bar h_{\sigma \gamma \epsilon} 
\nn \\ && 
- \bar{A} _{\mu} \nabla_{\nu}{\tau_{\rho}} \tau^{\nu} \tau^{\sigma}  h^{\mu \gamma} h^{\rho \epsilon} \bar h_{\sigma \gamma \epsilon} 
+ \frac12 \nabla_{\mu}{\bar{A} _{\nu}} \nabla_{\rho}{\bar{A} _{\sigma}}  h^{\mu \rho} h^{\nu \sigma} - \frac12 \nabla_{\mu}{\bar{A} _{\nu}} \nabla_{\rho}{\bar{A} _{\sigma}}  h^{\mu \sigma} h^{\nu \rho} 
\nn \\ && 
- \frac12 \chi^{2} \nabla_{\mu}{\bar{A} _{\nu}} \nabla_{\rho}{\tau_{\sigma}}  h^{\mu \sigma} h^{\nu \rho} + \frac12 \chi^{2} \nabla_{\mu}{\bar{A} _{\nu}} \nabla_{\rho}{\tau_{\sigma}}  h^{\mu \rho} h^{\nu \sigma}  
+ \frac12 \chi_{j} \nabla_{\mu}{\bar{A} _{\nu}}  f_{\rho \sigma}{}^{j} h^{\mu \rho} h^{\nu \sigma} 
\nn \\ && 
- \nabla_{\mu}{\bar{A} _{\nu}} \tau^{\rho}  h^{\mu \sigma} h^{\nu \gamma} \bar h_{\rho \sigma \gamma} 
+ \frac18 \chi^{2} \chi_{\xi} \chi^{\xi} \nabla_{\mu}{\tau_{\nu}} \nabla_{\rho}{\tau_{\sigma}}  h^{\mu \rho} h^{\nu \sigma} + \frac14 \chi^{2} \chi_{\xi} \nabla_{\mu}{\tau_{\nu}}  f_{\rho \sigma}{}^{\xi} h^{\mu \rho} h^{\nu \sigma} 
\nn \\ && 
- \frac18 \chi^{4} \nabla_{\mu}{\tau_{\nu}} \nabla_{\rho}{\tau_{\sigma}}  h^{\mu \sigma} h^{\nu \rho} - \frac12 \chi^{2} \nabla_{\mu}{\tau_{\nu}} \tau^{\rho}  h^{\mu \sigma} h^{\nu \gamma} \bar h_{\rho \sigma \gamma} 
+ \frac{1}{16} \chi_{k} \chi_{\xi}  f_{\mu \nu}{}^{k} f_{\rho \sigma}{}^{\xi} h^{\mu \rho} h^{\nu \sigma} 
\nn \\ && 
- \frac14 \chi_{j} \tau^{\mu}  f_{\nu \rho}{}^{j} h^{\nu \sigma} h^{\rho \gamma} \bar h_{\mu \sigma \gamma} + \frac14 \tau^{\mu} \tau^{\nu}  h^{\rho \sigma} h^{\gamma \epsilon} \bar h_{\mu \rho \gamma} \bar h_{\nu \sigma \epsilon} \, .
\eea
The previous expressions are given in a fully covariant form with respect to both diffeomorphisms and gauge transformations. Therefore, after consider the Carrollian ansatz of all the fields and taking the limit $w\rightarrow0$ the full bosonic sector of the Carrollian heterotic supergravity is 
\bea
S_{\rm het} = \int d^{10}x \, \Omega_{c} \, e^{-2 \varphi} (\hat R^{(2)} - 4 \partial_{\mu} \varphi \partial_{\nu} \varphi \tau^{\mu} \tau^{\nu} - \frac{1}{12} [{\bar H}^2]^{(2)} - \frac14 [\hat F_{\mu \nu i} \hat F^{\mu \nu i}]^{(2)}) \, .
\label{actionHET}
\eea

\subsection{Comments on the boost invariance of the action}
\label{Boost}
The action (\ref{actionHET}) is explicitly invariant under infinitesimal diffeomorphism and non-Abelian gauge transformations. If we further decompose,
\bea
h_{\mu \nu} = e_{\mu}{}^{a} \delta_{a b} e_{\nu}{}^{b} \, , \quad h^{\mu \nu} = e^{\mu}{}_{a} \delta^{a b} e^{\nu}{}_{b} \, ,
\eea
one can quickly verify that the action (\ref{actionHET}) is also invariant under spatial rotations given by \cite{Boost}
\bea
\delta_{\Lambda} e_{\mu}{}^{a} & = & \Lambda^{a}{}_{b} e_{\mu}{}^{b} \, , \\
\delta_{\Lambda} e^{\mu}{}_{a} & = & \Lambda_{a}{}^{b} e^{\mu}{}_{b} \, ,
\eea
where $\Lambda_{a b}$ is an arbitrary parameter and the rest of the fields are invariant under this symmetry. 

Finally, the action (\ref{actionHET}) is also invariant under boost transformations, but one needs to incorporate transformations for the b-field and a-field,
\bea
\delta \tau_{\mu} & = & e_{\mu}{}^{a} \lambda_{a} \, , \\
\delta_{\lambda} e^{\mu}{}_{a} & = & - \lambda_{a} \tau^{\mu} \, , \\
\delta_{\lambda} b_{\mu \nu} & = & - 2 e_{[\mu}{}^{a} A_{\nu]} \lambda_{a} \, , \\
\delta_{\lambda} a_{\mu}{}^{i} & = & - e_{\mu}{}^{a} \chi^{i} \lambda_{a} \, .
\eea
The rest of the fields ($\tau^{\mu}$, $e_{\mu}{}^{a}$, $A_{\mu}$, $\phi$ and $\chi^{i}$) are boost invariants. It is worthy to mention that after the redefinition $A_{\mu} \rightarrow \bar A_{\mu}$, the new $\bar A$-field inherits a boost transformation from the $a$-field. This indicates that the trivialization of the gauge Green-Schwarz mechanism of the 1-form induces a boost transformation on the redefined 1-form or, in other words, that the price to pay for trivializing the Green-Schwarz mechanism of the 1-form is a boost transformation.

\subsection{Black hole near-horizon applications and Carrollian string theory}

An interesting potential application of the Carrollian supergravity theories constructed in this work concerns the physics of strings propagating in the vicinity of black-hole horizons. It has recently been shown that the near-horizon dynamics of relativistic strings admits a natural Carrollian description, where the equations of motion reduce to those of electric and magnetic Carroll strings after performing the near-horizon expansion \cite{NearHorizon1}-\cite{NearHorizon2}. In particular, these analysis demonstrate that the Carrollian limit is not merely an abstract ultra-relativistic contraction, but rather provides an effective description of string propagation in physically relevant gravitational backgrounds.

If the correspondence between Carrollian worldsheet theories and Carrollian effective field theories holds, then the action derived in this work should provide the target-space description governing the background fields seen by Carrollian strings. In this sense, the present formulation may be regarded as the natural spacetime counterpart of the worldsheet constructions. In \cite{NearHorizon1}-\cite{NearHorizon2} the Schwarzschild geometry is expanded around the horizon, leading to a Carrollian decomposition of the target-space metric into longitudinal and transverse sectors. The resulting worldsheet dynamics splits into two distinct branches, corresponding to electric and magnetic Carroll strings, whose equations of motion are shown to reproduce the near-horizon limit of relativistic string propagation.

The effective theories constructed in the present work considerably enlarge this picture. Rather than considering only fixed background geometries, the Carrollian metric, the Kalb--Ramond field and the dilaton become fully dynamical degrees of freedom satisfying covariant field equations derived from an action principle. Consequently, the present formulation opens the possibility of studying backreacted Carrollian backgrounds, allowing one to investigate how string-induced matter sources may modify the near-horizon Carrollian geometry. Such a framework could provide a useful starting point for exploring black-hole solutions, horizon dynamics and Carrollian gravitational interactions directly within the effective theory. The extension to heterotic supergravity is particularly appealing from this perspective. The inclusion of non-Abelian gauge fields and the Green--Schwarz mechanism introduces the additional structures required for the low-energy description of heterotic strings.

\section{Equations of motion}
\label{EOM}
The dynamics of the Carrollian NS--NS and heterotic supergravity theory can be investigated from two complementary perspectives. One may either perform the ultra-relativistic expansion directly at the level of the relativistic equations of motion or derive the field equations from the finite Carrollian action through a variational principle. Since the Carrollian variables satisfy nonlinear constitutive relations, the equivalence between these two approaches is highly nontrivial. In this section we discuss both constructions and explain how they are expected to be related.  

\subsection{Expansion of the relativistic equations}

The equations of motion for the NS-NS/heterotic supergravity are given by
\bea
\Delta \hat g_{\mu \nu} & =&  \hat R_{\mu\nu}
+2\hat\nabla_\mu\hat\nabla_\nu\hat\phi
-\frac14
{\bar H}_{\mu\rho\sigma}
{\bar H}_{\nu}{}^{\rho\sigma}
-\frac12
\hat F_{\mu\rho i}
\hat F_{\nu}{}^{\rho i}
=0\, , \,\label{eomg}\\
\Delta \hat \phi & = & -2 \hat {\mathcal{L}} = 0 \, ,\label{eomphi}\\
\Delta \hat B_{\nu\rho} & =&  \frac{1}{2} \hat \nabla ^{\mu} \hat H_{\mu\nu\rho} - \hat \nabla ^{\mu} \hat \phi \hat H_{\mu\nu\rho} = 0\, \, , \ \ \ \ \ \label{eomb}\\
\Delta \hat A_{\mu}{}^{i} & = & \frac{1}{2} \hat H_{\mu\nu\rho} \hat F^{\nu\rho i} + \hat A_{\rho}{}^{i}\Delta \hat B^{\rho}{}_{\mu} - \hat \nabla^{\nu} \hat F_{\mu\nu}{}^{i} + 2 \hat F_{\mu\nu}{}^{i} \hat \nabla^{\nu} \hat \phi = 0 \, ,\label{eoma} 
\eea
where $\hat A_{\mu}{}^{i}=0$ must be imposed to obtain the NS-NS case.

After applying the Carrollian ansatz (\ref{Metric}), (\ref{B}),(\ref{scaling}) and (\ref{A}) on every field we find the following equations:
\paragraph{Metric equations}
\bea
\label{EOMinicial}
\Delta \hat g_{\mu \nu}^{(-4)} & = & \hat R^{(-4)}_{\mu \nu}=0 \, , \\
\Delta \hat g_{\mu \nu}^{(-2)} & = & \hat R^{(-2)}_{\mu \nu} + 2 (\hat \nabla_{\mu}\hat \nabla_{\nu}\hat{\phi})^{(-2)}=0 \, , \\
\Delta \hat g_{\mu \nu}^{(0)} & = & \hat R^{(0)}_{\mu \nu} + 2 (\hat \nabla_{\mu}\hat \nabla_{\nu}\hat{\phi})^{(0)} \nn \\ && - \frac14 \hat{H}_{\mu \rho \sigma} \hat{H}_{\nu \gamma \epsilon} h^{\rho \gamma} h^{\sigma \epsilon}    - \frac12 \hat{F}_{\mu \rho}^{i} \hat{F}_{\nu \sigma i} h^{\rho \sigma}=0 \, , \\
\Delta \hat g_{\mu \nu}^{(2)} & = & \hat R^{(2)}_{\mu \nu} + 2 (\hat \nabla_{\mu}\hat \nabla_{\nu}\hat{\phi})^{(2)} + \frac14 \hat{H}_{\mu \rho \sigma} \hat{H}_{\nu \gamma \epsilon} \tau^{\sigma} \tau^{\epsilon} h^{\rho \gamma} \nn \\ && + \frac14 \hat{H}_{\mu \rho \sigma} \hat{H}_{\nu \gamma \epsilon} \tau^{\rho} \tau^{\gamma} h^{\sigma \epsilon} + \frac12 \hat{F}_{\mu \rho}^{i} \hat{F}_{\nu \sigma i} \tau^{\rho} \tau^{\sigma}=0 \, , \\
\Delta \hat g_{\mu \nu}^{(4)} & = & - \frac14 \hat{H}_{\mu \rho \sigma} \hat{H}_{\nu \gamma \epsilon} \tau^{\rho} \tau^{\sigma} \tau^{\gamma} \tau^{\epsilon}=0 \, .
\eea

\paragraph{Dilaton equations}
\bea
\Delta \hat{\phi}^{(-2)} & = & \hat R^{(-2)} = 0 \, , \\
\Delta \hat{\phi}^{(0)} & = & \hat R^{(0)} + 4 \partial_{\mu}{\hat{\phi}} \partial_{\nu}{\hat{\phi}} h^{\mu \nu} - \frac{1}{12} \hat{H}_{\mu \nu \rho} \hat{H}_{\sigma \gamma \epsilon} h^{\mu \sigma} h^{\nu \gamma} h^{\rho \epsilon} \nn \\ &&  - \frac14 \hat{F}_{\mu \nu}^{i} \hat{F}_{\rho \sigma i} h^{\mu \rho} h^{\nu \sigma} = 0 \, , \\
\Delta \hat{\phi}^{(2)} & = & \hat R^{(2)} - 4 \partial_{\mu}{\hat{\phi}} \partial_{\nu}{\hat{\phi}} \tau^{\mu} \tau^{\nu} + \frac{1}{12} \hat{H}_{\mu \nu \rho} \hat{H}_{\sigma \gamma \epsilon} \tau^{\rho} \tau^{\epsilon} h^{\mu \sigma} h^{\nu \gamma} \nn \\ && + \frac{1}{12} \hat{H}_{\mu \nu \rho} \hat{H}_{\sigma \gamma \epsilon} \tau^{\nu} \tau^{\gamma} h^{\mu \sigma} h^{\rho \epsilon} + \frac{1}{12} \hat{H}_{\mu \nu \rho} \hat{H}_{\sigma \gamma \epsilon} \tau^{\mu} \tau^{\sigma} h^{\nu \gamma} h^{\rho \epsilon} \nn \\ && + \frac14 \hat{F}_{\mu \nu}^{i} \hat{F}_{\rho \sigma i} \tau^{\nu} \tau^{\sigma} h^{\mu \rho} + \frac14 \hat{F}_{\mu \nu}^{i} \hat{F}_{\rho \sigma i} \tau^{\mu} \tau^{\rho} h^{\nu \sigma} = 0 \, , \\
\Delta \hat{\phi}^{(4)} & = & - \frac{1}{12} \hat{H}_{\mu \nu \rho} \hat{H}_{\sigma \gamma \epsilon} \tau^{\nu} \tau^{\rho} \tau^{\gamma} \tau^{\epsilon} h^{\mu \sigma} - \frac{1}{12} \hat{H}_{\mu \nu \rho} \hat{H}_{\sigma \gamma \epsilon} \tau^{\mu} \tau^{\rho} \tau^{\sigma} \tau^{\epsilon} h^{\nu \gamma} \nn \\ && - \frac{1}{12} \hat{H}_{\mu \nu \rho} \hat{H}_{\sigma \gamma \epsilon} \tau^{\mu} \tau^{\nu} \tau^{\sigma} \tau^{\gamma} h^{\rho \epsilon} - \frac14 \hat{F}_{\mu \nu}^{i} \hat{F}_{\rho \sigma i} \tau^{\mu} \tau^{\nu} \tau^{\rho} \tau^{\sigma}=0 \, , \\
\Delta \hat{\phi}^{(6)} & = & \frac{1}{12} \hat{H}_{\mu \nu \rho} \hat{H}_{\sigma \gamma \epsilon} \tau^{\mu} \tau^{\nu} \tau^{\rho} \tau^{\sigma} \tau^{\gamma} \tau^{\epsilon} = 0 \, .
\eea

\paragraph{$\hat B$-field equations}
\bea
\Delta \hat B_{\mu \nu}^{(-2)} & = & \frac12 (\hat \nabla_{\rho}\hat{H}_{ \sigma \mu \nu})^{(-2)} h^{\rho \sigma} =0 \, , \\
\Delta \hat B_{\mu \nu}^{(0)} & = & \frac12 (\hat \nabla_{\rho}\hat{H}_{ \sigma \mu \nu})^{(0)} h^{\rho \sigma} - \frac12 (\hat \nabla_{\rho}\hat{H}_{\sigma \mu \nu})^{(-2)} \tau^{\rho} \tau^{\sigma} + \hat{H}_{\rho \mu \nu} \partial_{\sigma}{\varphi} h^{\rho \sigma} = 0 \, , \\
\Delta \hat B_{\mu \nu}^{(2)} & = & \frac12 (\hat \nabla_{\rho} \hat{H}_{\sigma \mu \nu})^{(2)} h^{\rho \sigma} - \frac12 (\hat \nabla_{\rho}\hat{H}_{\sigma \mu \nu})^{(0)} \tau^{\rho} \tau^{\sigma} - \hat{H}_{\rho \mu \nu} \partial_{\sigma}{\varphi} \tau^{\rho} \tau^{\sigma} = 0 , \\
\Delta \hat B_{\mu \nu}^{(4)} & = & - \frac12 (\hat \nabla_{\rho} \hat{H}_{ \sigma \mu \nu})^{(2)} \tau^{\rho} \tau^{\sigma} = 0 \, .
\eea

\paragraph{$\hat A$-field equations}
\bea
\Delta \hat A_{\mu}^{i(-2)} & = & - \hat A_{\nu}^{i} (\Delta \hat B_{\mu \rho})^{(-2)} h^{\nu \rho} - (\hat \nabla_{\nu} \hat F_{\mu \rho}^{i})^{(-2)} h^{\nu \rho} =0 \, , \\
\Delta \hat A_{\mu}^{i(0)} & = & \frac12 \hat F_{\nu \rho}{}^{i} \hat H_{\mu \sigma \gamma} h^{\nu \sigma} h^{\rho \gamma} - \hat A_{\nu}^{i} (\Delta \hat B_{\mu \rho})^{(0)} h^{\nu \rho} + \hat A_{\nu}^{i} (\Delta \hat B_{\mu \rho})^{(-2)} \tau^{\nu} \tau^{\rho} \nn \\ && + (\hat \nabla_{\nu} \hat F_{\mu \rho}^{i})^{(-2)} \tau^{\nu} \tau^{\rho} + 2 \hat F_{\mu \nu}^{i} \partial_{\rho}{\varphi} h^{\nu \rho} = 0 \, , \\
\Delta \hat A_{\mu}^{i(2)} & = & - \frac12 \hat F_{\nu \rho}{}^{i} \hat H_{\mu \sigma \gamma} \tau^{\rho} \tau^{\gamma} h^{\nu \sigma} - \frac12 \hat F_{\nu \rho}{}^{i} H_{\mu \sigma \gamma} \tau^{\nu} \tau^{\sigma} h^{\rho \gamma} - \hat A_{\nu}^{i} (\Delta \hat B_{\mu \rho})^{(2)} h^{\nu \rho} \nn \\ && + \hat A_{\nu}^{i} (\Delta \hat B_{\mu \rho})^{(0)} \tau^{\nu} \tau^{\rho} - (\hat \nabla_{\nu}\hat F_{\mu \rho}^{i})^{(2)} h^{\nu \rho} - (\hat \nabla_{\nu} \hat F_{\mu \rho}^{i})^{(0)} h^{\nu \rho} + (\hat \nabla_{\nu} \hat F_{\mu \rho}^{i})^{(0)} \tau^{\nu} \tau^{\rho} \nn \\ && - 2 \hat F_{\mu \nu}^{i} \partial_{\rho}{\varphi} \tau^{\nu} \tau^{\rho}= 0 , \\
\Delta \hat A_{\mu}^{i(4)} & = & \frac12 \hat F_{\nu \rho}{}^{i} \hat H_{\mu \sigma \gamma} \tau^{\nu} \tau^{\rho} \tau^{\sigma} \tau^{\gamma} + \hat A_{\nu}^{i} (\Delta \hat B_{\mu \rho})^{(2)} \tau^{\nu} \tau^{\rho} + (\hat \nabla_{\nu}\hat F_{\mu \rho}^{i})^{(2)} \tau^{\nu} \tau^{\rho} = 0 \, .
\label{EOMfinal}
\eea
We provide the explicit expansion for the Carrollian dilaton gravity case \footnote{It is interesting to observe that, in the non-relativistic limit of NS-NS \cite{NSNS} and heterotic supergravity \cite{BandR}-\cite{LO}, one cannot consistently set $\hat H_{\mu \nu \rho}=0$, since the contribution of this field to the action is required to cancel divergences. This is in contrast to the Carrollian case.} in the appendix \ref{App}. While the previous equations respect the Carrollian decomposition, new geometrical constraints should be imposed to compatibilize this method with the explicit variation of the finite Lagrangian.
 
\subsection{Equations of motion through a variational principle}

In the previous subsection we obtained the Carrollian equations of motion by expanding the relativistic field equations around the ultra-relativistic limit. In this section we derive the dynamics directly from the Carrollian action through a variational principle. We illustrate this procedure for the Carrollian dilaton-gravity sector, obtained by consistently truncating the antisymmetric tensor and gauge fields,
\begin{equation}
A_\mu=b_{\mu\nu}=a_\mu{}^i=\chi^i=0.
\end{equation}
Therefore, the fundamental fields are $\{\tau_\mu,\tau^\mu,h_{\mu\nu},h^{\mu\nu},\varphi\}$, which satisfy the Carrollian constitutive relations (\ref{cr1})-(\ref{cr3}).
Rather than solving these constraints explicitly before varying the action, we shall regard all fundamental fields as independent variables. The constitutive relations are then imposed through Lagrange multipliers. This procedure has the advantage of treating all geometric fields democratically while avoiding the complications associated with constrained variations.

The action to be varied is therefore
\begin{equation}
S=S_{\mathrm{CG}}
+S_{\mathrm{LM}},
\end{equation}
where $S_{\mathrm{CG}}$ denotes the Carrollian dilaton-gravity action obtained in (\ref{actionNSNS}) with $\hat H_{\mu \nu \rho}=0$ and
\begin{align}
S_{\mathrm{LM}}
=\int d^Dx\,\Omega\,
\Big[
&
\lambda\,(\tau_\mu\tau^\mu-1)
+\lambda_\nu\,(\tau_\mu h^{\mu\nu})
+{\lambda}^{\mu}(\tau^\nu h_{\mu\nu})
\nonumber\\
&
+\lambda^\mu{}_\nu
\left(
h_{\mu\rho}h^{\rho\nu}
+\tau_\mu\tau^\nu
-\delta_\mu^\nu
\right)
\Big].
\label{Lagrange}
\end{align}
Here $\lambda$, $\lambda_\mu$, ${\lambda}^{\mu}$ and $\lambda^\mu{}_\nu$ are independent Lagrange multiplier fields enforcing the Carrollian relations and $\Omega = \Omega_{c} e^{-2\varphi}$ is the Carrollian measure. The total variation of the action can then be written as
\begin{equation}
\delta S
=
\delta_{\tau_\mu}S
+\delta_{\tau^\mu}S
+\delta_{h_{\mu\nu}}S
+\delta_{h^{\mu\nu}}S
+\delta_\varphi S
+\delta S_{\mathrm{LM}},
\end{equation}
where each contribution is treated independently. For computing the variation we need the relation between the measure and the trace of the connection, given by
\bea
\partial_{\nu}(\ln \Omega_{c}) = \Gamma_{\mu \nu}^{\mu} = \tau^{\mu} \partial_{\nu} \tau_{\mu} + \frac12 h^{\rho \sigma} \partial_{\nu} h_{\rho \sigma} \, .
\eea
From the previous expression we learn that
\bea
\delta(\Omega_{c})  = \Omega_{c} ( \tau^{\mu} \delta \tau_{\mu} + \frac12 h^{\rho \sigma} \delta h_{\rho \sigma}) \, .
\eea
The variation of the Carrollian dilaton gravity action with respect to the fundamental fields is given in the Appendix \ref{App}. Finally, varying the Lagrange multipliers reproduces the Carrollian constitutive relations, whereas the variations of the geometric fields receive additional contributions proportional to the Lagrange multipliers. Eliminating the latter yields the independent Carrollian equations of motion. In order to have a consistent truncation, these equations should coincide with those obtained from the ultra-relativistic expansion of the relativistic equations of motion (after imposing geometric constraints).

\section{Discussion}
\label{Discussion}

\subsection{Comparison with NS--NS and heterotic non-relativistic supergravity}

The construction presented in this work shares several features with the non-relativistic formulation of NS--NS supergravity developed in \cite{NSNS}. The main similarity is the use of the dilaton to control the divergences arising from the determinant of the metric. In our formulation, this is the key ingredient that makes it possible to consistently couple Carrollian gravity to the dilaton and Kalb--Ramond sectors, whose individual contributions all diverge in the ultra-relativistic limit. While, in the non-relativistic construction, there is a non-trivial cancellation between the Ricci scalar and the $H^2$ term, no such cancellation occurs in the Carrollian limit. As a result, these divergences must instead be compensated by the scaling of the measure. Regarding the dynamics, the non-relativistic limit of NS-NS supergravity contains an emergent symmetry, the dilatation symmetry, which effectively explains the absence of the Poisson equation and makes the introduction of the torsion constraints more transparent. In the Carrollian case, we also expect the emergence of new symmetries \cite{V2citation1}-\cite{V2citation7} from the compatibility between the variational principle and the expansion of the relativistic equations of motion. 

Turning to the heterotic case, two non-relativistic formulations have recently been constructed. On the one hand, there is the formulation \cite{BandR}, in which the expansion of the vielbein contains the gauge fields and the Green--Schwarz mechanism is unambiguously realized. On the other hand, there is the formulation \cite{LO}, based on a finite uplift of the generalized metric in Double Field Theory \footnote{There is a clear connection between non-Lorentzian geometries and non-riemannian Double Field Theory \cite{NRDFT1}-\cite{NRDFT5}. It would be very interesting to compare the NS-NS construction of this paper and the non-Riemannian parametrization of \cite{NRDFT4} to inspect if they are the same theory after a shifting in the connection. Similarly, it would be very interesting to study the uplift of our heterotic extension to the double geometry.} \cite{DFT1}-\cite{DFT4} (see \cite{ReviewDFT1}-\cite{ReviewDFT3} for reviews). In this approach, the Green--Schwarz mechanism can be trivialized \cite{Eheterotic}, yielding a gauge-invariant redefinition of the $b$-field. In the Carrollian limit, the situation is intermediate: the Green--Schwarz transformation of the 1-form field can be trivialized through the field redefinition (\ref{barA}), whereas the transformation of the $b$-field cannot. In this sense, the Carrollian limit constructed in the present work interpolates between the two non-relativistic formulations with respect to the realization of the Green--Schwarz mechanism.

\subsection{$\alpha'$-corrections}

Regarding $\alpha'$ corrections, recent progress has also been made in the non-relativistic limit for both bosonic supergravity \cite{EL} and heterotic supergravity \cite{HeteroticAP1}-\cite{HeteroticAP2}. Even after performing the rescaling $\alpha'\rightarrow \frac{\alpha'_{\rm NR}}{c^2}$, the non-relativistic ansatz still produces divergences of order $c^2$ in the bosonic theory, originating from $c^4$ contributions to the curvature invariants. Divergences of the analogous order ($w^4$) could instead be controlled within the Carrollian ansatz (plus a rescaling $\alpha'\rightarrow \frac{\alpha'_{\rm c}}{w^2}$) thanks to the additional $\frac{1}{w^2}$ factor provided by the measure. While it remains unclear whether higher-order divergences, such as those of order $w^6$ or beyond, may arise from curvature-squared terms, for instance from a $\mathrm{Riem}^2$ contribution or $(H^2)^2$. For example, the potential divergences from $\mathrm{Riem}^2$ are given by 
\bea
[\hat R^{\mu}{}_{\nu \rho \sigma} \hat R^{\alpha}{}_{\beta \gamma \delta} \hat g_{\mu \alpha} \hat g^{\nu \beta} g^{\rho \gamma} g^{\sigma \delta}]^{(10)} =  \hat R^{(2)\mu}{}_{\nu \rho \sigma} \hat R^{(2)\gamma}{}_{\epsilon \delta \lambda} \tau^{\nu} \tau^{\rho} \tau^{\sigma} \tau^{\epsilon} \tau^{\delta} \tau^{\lambda} h_{\mu \gamma} \, ,
\eea
\bea
&& [\hat R^{\mu}{}_{\nu \rho \sigma} \hat R^{\alpha}{}_{\beta \gamma \delta} \hat g_{\mu \alpha} \hat g^{\nu \beta} g^{\rho \gamma} g^{\sigma \delta}]^{(8)} =  \hat R^{(2)\mu}{}_{\nu \rho \sigma} \hat R^{(2)\gamma}{}_{\epsilon \delta \lambda} \tau^{\rho} \tau^{\sigma} \tau^{\delta} \tau^{\lambda}  h_{\mu \gamma} h^{\nu \epsilon} \nn \\ && + \hat R^{(2)\mu}{}_{\nu \rho \sigma} \hat R^{(2)\gamma}{}_{\epsilon \delta \lambda} \tau^{\nu} \tau^{\sigma} \tau^{\epsilon} \tau^{\lambda}  h_{\mu \gamma} h^{\rho \delta}  + \hat R^{(2)\mu}{}_{\nu \rho \sigma} \hat R^{(2)\gamma}{}_{\epsilon \delta \lambda} \tau^{\nu} \tau^{\rho} \tau^{\epsilon} \tau^{\delta}  h_{\mu \gamma} h^{\sigma \lambda} \nn \\ && + \hat R^{(2)\mu}{}_{\nu \rho \sigma} \hat R^{(2)\gamma}{}_{\epsilon \delta \lambda} \tau_{\mu} \tau_{\gamma} \tau^{\nu} \tau^{\rho} \tau^{\sigma} \tau^{\epsilon} \tau^{\delta} \tau^{\lambda} + 2 \hat R^{(0)\mu}{}_{\nu \rho \sigma} \hat R^{(2)\gamma}{}_{\epsilon \delta \lambda} \tau^{\nu} \tau^{\rho} \tau^{\sigma} \tau^{\epsilon} \tau^{\delta} \tau^{\lambda}  h_{\mu \gamma} \, ,
\eea
\bea
&& [\hat R^{\mu}{}_{\nu \rho \sigma} \hat R^{\alpha}{}_{\beta \gamma \delta} \hat g_{\mu \alpha} \hat g^{\nu \beta} g^{\rho \gamma} g^{\sigma \delta}]^{(6)} = \hat R^{(2)\mu}{}_{\nu \rho \sigma} \hat R^{(2)\gamma}{}_{\epsilon \delta \lambda} \tau^{\sigma} \tau^{\lambda}  h_{\mu \gamma} h^{\nu \epsilon} h^{\rho \delta} \nn \\ && + \hat R^{(2)\mu}{}_{\nu \rho \sigma} \hat R^{(2)\gamma}{}_{\epsilon \delta \lambda} \tau^{\rho} \tau^{\delta}  h_{\mu \gamma} h^{\nu \epsilon} h^{\sigma \lambda} + \hat R^{(2)\mu}{}_{\nu \rho \sigma} \hat R^{(2)\gamma}{}_{\epsilon \delta \lambda} \tau_{\mu} \tau_{\gamma} \tau^{\rho} \tau^{\sigma} \tau^{\delta} \tau^{\lambda}  h^{\nu \epsilon} \nn \\ && + \hat R^{(2)\mu}{}_{\nu \rho \sigma} \hat R^{(2)\gamma}{}_{\epsilon \delta \lambda} \tau^{\nu} \tau^{\epsilon}  h_{\mu \gamma} h^{\rho \delta} h^{\sigma \lambda}  + \hat R^{(2)\mu}{}_{\nu \rho \sigma} \hat R^{(2)\gamma}{}_{\epsilon \delta \lambda} \tau_{\mu} \tau_{\gamma} \tau^{\nu} \tau^{\sigma} \tau^{\epsilon} \tau^{\lambda}  h^{\rho \delta} \nn \\ && + \hat R^{(2)\mu}{}_{\nu \rho \sigma} \hat R^{(2)\gamma}{}_{\epsilon \delta \lambda} \tau_{\mu} \tau_{\gamma} \tau^{\nu} \tau^{\rho} \tau^{\epsilon} \tau^{\delta}  h^{\sigma \lambda} + 2 \hat R^{(0)\mu}{}_{\nu \rho \sigma} \hat R^{(2)\gamma}{}_{\epsilon \delta \lambda} \tau^{\rho} \tau^{\sigma} \tau^{\delta} \tau^{\lambda}  h_{\mu \gamma} h^{\nu \epsilon} \nn \\ && + 2 \hat R^{(0)\mu}{}_{\nu \rho \sigma} \hat R^{(2)\gamma}{}_{\epsilon \delta \lambda} \tau^{\nu} \tau^{\sigma} \tau^{\epsilon} \tau^{\lambda}  h_{\mu \gamma} h^{\rho \delta} + 2 \hat R^{(0)\mu}{}_{\nu \rho \sigma} \hat R^{(2)\gamma}{}_{\epsilon \delta \lambda} \tau^{\nu} \tau^{\rho} \tau^{\epsilon} \tau^{\delta}  h_{\mu \gamma} h^{\sigma \lambda} \nn \\ && + 2 R^{(0)\mu}{}_{\nu \rho \sigma} \hat R^{(2)\gamma}{}_{\epsilon \delta \lambda} \tau_{\mu} \tau_{\gamma} \tau^{\nu} \tau^{\rho} \tau^{\sigma} \tau^{\epsilon} \tau^{\delta} \tau^{\lambda} + 2 \hat R^{(2)\mu}{}_{\nu \rho \sigma} \hat R^{(-2)\gamma}{}_{\epsilon \delta \lambda} \tau^{\nu} \tau^{\rho} \tau^{\sigma} \tau^{\epsilon} \tau^{\delta} \tau^{\lambda}  h_{\mu \gamma} \nn \\ && + \hat R^{(0)\mu}{}_{\nu \rho \sigma} \hat R^{(0)\gamma}{}_{\epsilon \delta \lambda} \tau^{\nu} \tau^{\rho} \tau^{\sigma} \tau^{\epsilon} \tau^{\delta} \tau^{\lambda}  h_{\mu \gamma} \, . 
\eea
After a long but straighforward computation we verify that 
$[\hat R^{\mu}{}_{\nu \rho \sigma} \hat R^{\alpha}{}_{\beta \gamma \delta} \hat g_{\mu \alpha} \hat g^{\nu \beta} g^{\rho \gamma} g^{\sigma \delta}]^{(10)}= [\hat R^{\mu}{}_{\nu \rho \sigma} \hat R^{\alpha}{}_{\beta \gamma \delta} \hat g_{\mu \alpha} \hat g^{\nu \beta} g^{\rho \gamma} g^{\sigma \delta}]^{(8)}= [\hat R^{\mu}{}_{\nu \rho \sigma} \hat R^{\alpha}{}_{\beta \gamma \delta} \hat g_{\mu \alpha} \hat g^{\nu \beta} g^{\rho \gamma} g^{\sigma \delta}]^{(6)}=0$.
This verification implies that the four-derivative bosonic (heterotic) $\alpha'$-corrections with $\hat H_{\mu \nu \rho}=0$ ($\bar H_{\mu \nu \rho}=0$) are finite, allow further explorations of higher-derivative Carrollian gravities.

\subsection{Connection to the worldsheet formalism}

An important question concerns the relation between the Carrollian NS--NS theory constructed in this work and recent developments on Carrollian string theory. While our analysis has been entirely performed from the spacetime perspective by taking the ultra-relativistic limit of the relativistic NS--NS supergravity action, a complementary worldsheet description has recently been proposed in \cite{FigueroaNew}, where a Carrollian bosonic string is obtained directly from the Polyakov action through an ultra-relativistic limit on the worldsheet. Although these two constructions are formulated in different languages, they exhibit several remarkable similarities, suggesting that they may correspond to different descriptions of the same underlying theory.

The most immediate observation concerns the field content. The universal bosonic sector of relativistic string theory is described by the metric, the Kalb--Ramond 2-form and the dilaton. After performing the Carrollian decomposition, these fields give rise precisely to the fundamental variables of the present construction, without introducing additional geometric degrees of freedom. In particular, the 1-form $A_\mu$ originates from the mixed temporal-spatial components of the Kalb--Ramond field, while the spatial 2-form $b_{\mu\nu}$ and the dilaton survive as independent Carrollian fields. Consequently, the field content obtained from the spacetime contraction coincides with the natural decomposition of the universal massless bosonic string multiplet.

A second point of agreement concerns the critical dimension. The BRST analysis of the Carrollian bosonic string performed in \cite{FigueroaNew} shows that quantum consistency requires a $26$-dimensional target space, exactly as in relativistic bosonic string theory. Since the Carrollian NS--NS action derived in the present work is valid in arbitrary dimension, it admits a consistent specialization to $D=26$. This observation naturally raises the possibility that the theory constructed here provides the low-energy spacetime description associated with the critical Carrollian bosonic string.

The analogy with relativistic string theory further supports this interpretation. In the relativistic case, the Polyakov sigma model and the NS--NS supergravity action constitute complementary descriptions of the same physics: the vanishing of the worldsheet beta functions determines the equations of motion of the spacetime effective theory. The present construction reproduces the complete interacting Carrollian dynamics of the metric, Kalb--Ramond field and dilaton, suggesting that it should play an analogous role for Carrollian string theory. From this perspective, the Carrollian action derived in this work may be regarded as a natural candidate for the effective field theory governing the massless bosonic sector of the Carrollian string.

At present, however, this interpretation should be regarded as conjectural. The worldsheet formulation of \cite{FigueroaNew} has been developed in flat target space, and no computation of the beta functions for a general Carrollian sigma model is currently available (similarly for the recent findings in \cite{V2citation1}-\cite{V2citation7}, where a new Carrollian gauge symmetry was identified and analyzed). Consequently, there is not yet a direct derivation of the spacetime field equations from worldsheet conformal invariance analogous to the relativistic case. Establishing such a correspondence would require coupling the Carrollian worldsheet theory to the background fields and computing the corresponding quantum consistency conditions.

\section{Conclusions}
\label{Conclusions}

In this work we have constructed the Carrollian limit of the bosonic NS--NS and heterotic supergravity theories by performing a systematic ultra-relativistic expansion of the corresponding relativistic actions. The resulting formulations provide finite, covariant theories written entirely in terms of Carrollian geometric variables, extending previous studies of Carrollian gravity to include the complete bosonic field content of string effective theories.

For the NS--NS sector, we introduced a Carrollian decomposition of the metric, the Kalb--Ramond field and the dilaton that yields a finite action after taking the limit $w\rightarrow\infty$. A key ingredient of the construction is the scaling of the dilaton given in (\ref{scaling}), whose contribution to the measure compensates the divergences generated by both the Ricci scalar and the 3-form field strength. In contrast with the non-relativistic limit, where the gravitational and antisymmetric tensor sectors exhibit non-trivial cancellations, the Carrollian formulation requires the measure itself to regulate the divergent contributions.

The geometric structure underlying the theory was also investigated in detail. Starting from the expansion of the relativistic Levi--Civita connection, we constructed the intrinsic Carrollian affine connection together with its associated curvature and effective non-metricity tensors (\ref{metricity1})-(\ref{metricity4}). Rather than introducing these quantities independently, they arise uniquely from the relativistic contraction, providing a covariant geometric framework compatible with the Carrollian constitutive relations. This construction allows the finite NS-NS action (\ref{actionNSNS}) to be written entirely in terms of intrinsic Carrollian geometric objects and in a covariant way with respect to infinitesimal diffeomorphisms.

The analysis was subsequently extended to heterotic supergravity by incorporating the non-Abelian gauge sector together with the Green--Schwarz mechanism. The ultra-relativistic limit produces a finite Carrollian Yang--Mills theory consistently coupled to the gravitational, dilaton and Kalb--Ramond sectors (the explicit action is given in (\ref{actionHET})). An interesting feature of the construction is that the Green--Schwarz transformation of the Carrollian 1-form, $A_{\mu}$, can be eliminated through the field redefinition (\ref{barA}), whereas the corresponding transformation of the 2-form $b_{\mu \nu}$ remains non-ambiguous. In this sense, the heterotic Carrollian theory exhibits properties that interpolate between the two non-relativistic limits of heterotic supergravity obtained in the literature. Particularly, this field redefinition induces new boost transformations on the (redefined) 1-form, as we discussed in section \ref{Boost}.

To further investigate the dynamics of the Carrollian theory, we analyzed the equations of motion from two complementary perspectives. On the one hand, the relativistic field equations were expanded directly in the Carrollian limit. On the other hand, we varied the Carrollian action by treating the fundamental geometric fields as independent variables and imposing the Carrollian constitutive relations through Lagrange multipliers, restricting the presentation to the Carrollian dilaton gravity sector. These two approaches are expected to provide equivalent descriptions of the dynamics. However, establishing this correspondence explicitly requires solving the constraint equations associated with the Lagrange multipliers given in (\ref{Lagrange}), a task that appears to be highly non-trivial in the general case and may depend on the class of solutions under consideration. It is also possible that additional or emergent symmetries play a role in relating the two formulations. Due to the complexity of the constraints even for a simplified scenario, such as dilaton gravity (the variations for this case are provided in the appendix \ref{App}), we leave a detailed comparison between the two approaches for future work \cite{BLP2}.

Finally, we discussed several interpretations of the resulting theories. In particular, the field content and geometric structure of the Carrollian NS--NS sector suggest that it may provide a natural candidate for the effective spacetime description associated with Carrollian bosonic string theory in the critical dimension. Regarding $\alpha'$-corrections, we explicitly checked that the $\hat{\rm Riem}^2$ contribution of both bosonic and heterotic supergravity is finite, when one rescales $\alpha'\rightarrow \frac{\alpha'_c}{w^2}$ (there are no $\mathcal{O}(w^6)$ divergences when $\hat H_{\mu \nu \rho}=0$). This opens a bright possibility to explore solutions for more general setups. We also highlighted potential applications to string propagation near black-hole horizons, where recent worldsheet analyses indicate that Carrollian dynamics emerge naturally in the near-horizon regime. By complementing recent developments in non-relativistic effective NS-NS and heterotic supergravity, our results contribute towards the understanding of the role of non-Lorentzian geometry in quantum gravity.

\subsection*{Acknowledgements}
The work of R.B is supported by Vicerrectoría de Investigación y Doctorados, Universidad San Sebastián, through the postdoctoral project USS-FIN-26-PINX-03. The work of E.L is supported by the SONATA BIS grant 2021/42/E/ST2/00304 from the National Science Centre (NCN), Poland. The work of S.P.L was partially supported by DGAPA-UNAM grant IN116823.

\appendix
\section{Variation of the Carrollian dilaton gravity}
\label{App}
The variation of the Carrollian dilaton gravity action with respect to the dilaton is
\bea
\frac{\delta S}{\delta \varphi} = \Omega_{c} e^{-2\varphi} L + 8 \partial_{\mu}(\Omega_{c} e^{-2\phi} \partial_{\nu}\varphi \tau^{\mu} \tau^{\nu}) = 0\, .
\eea
The variation with respect to the clock 1-form gives
\bea
\frac{\delta S}{\delta \tau_{\nu}} & = & \Omega_{c} \tau^{\nu} e^{-2\varphi}L + \Omega_{c} e^{-2\varphi}[-4\partial_{\mu} \tau^{\mu} \partial_{\sigma}\tau^{\nu} \tau^{\sigma} 
\nn \\ &&
+ 9 \partial_{\mu} \tau^{\nu} \partial_{\rho}\tau^{\sigma} \tau_{\sigma} \tau^{\mu} \tau^{\rho} + 2 \partial_{\mu} \tau_{\sigma} \partial_{\rho} \tau^{\nu} \tau^{\mu} \tau^{\sigma} \tau^{\rho} 
\nn \\ &&
- 2 \partial_{\mu} \tau^{\nu} \partial_{\rho}h_{\sigma \gamma} \tau^{\mu} \tau^{\rho} h^{\sigma \gamma} - 2 \partial_{\mu} \tau^{\rho} \partial_{\rho}\tau^{\nu} \tau^{\mu}] 
\nn \\ &&
+ 2 \partial_{\mu}(\partial_{\rho} \tau^{\rho} \tau^{\mu} \tau^{\nu} \Omega_{c} e^{2\varphi}) 
- 2 \partial_{\mu}(\partial_{\rho} \tau^{\sigma} \tau_{\sigma} \tau^{\mu} \tau^{\nu} \tau^{\rho} \Omega_{c} e^{2\varphi})
\nn \\ &&
+ \partial_{\mu}(\partial_{\rho} h_{\sigma \gamma} \tau^{\mu} \tau^{\nu} \tau^{\rho} h^{\sigma \gamma} \Omega_{c} e^{2\varphi}) = 0 \, ,
\eea
while varying the dual clock vector gives
\bea
\frac{\delta S}{\delta \tau^{\alpha}} & = & \Omega_{c} e^{-2\varphi} [-2 \partial_{\alpha} \tau^{\nu} \partial_{\rho}h_{\nu \sigma} h^{\rho \sigma} - 4 \partial_{\mu}\tau^{\mu} \partial_{\alpha}\tau^{\rho} \tau_{\rho} + 9 \partial_{\alpha} \tau^{\nu} \partial_{\rho}\tau^{\sigma} \tau_{\nu} \tau_{\sigma} \tau^{\rho} 
\nn \\ && 
- 2 \partial_{\alpha}\tau_{\nu} \partial_{\rho} \tau^{\rho} \tau^{\nu} - 2 \partial_{\mu} \tau_{\alpha} \partial_{\rho} \tau^{\rho} \tau^{\mu}
+ 2 \partial_{\alpha}\tau_{\nu} \partial_{\rho}\tau^{\sigma} \tau_{\sigma} \tau^{\nu} \tau^{\rho} + 2 \partial_{\mu} \tau_{\alpha} \partial_{\rho}\tau^{\sigma} \tau_{\sigma} \tau^{\mu} \tau^{\rho}
\nn \\ &&
+ 2 \partial_{\mu} \tau_{\nu} \partial_{\alpha} \tau^{\sigma} \tau_{\sigma} \tau^{\mu} \tau^{\nu} + \partial_{\mu} \tau^{\mu} \partial_{\alpha} h_{\rho \sigma} h^{\rho \sigma}
-2 \partial_{\alpha} \tau^{\nu} \partial_{\rho}h_{\sigma \gamma} \tau_{\nu} \tau^{\rho} h^{\sigma \gamma} - 2 \partial_{\mu} \tau^{\nu} \partial_{\alpha}h_{\sigma \gamma} \tau_{\nu} \tau^{\mu} h^{\sigma \gamma}
\nn \\ &&
+ \partial_{\alpha} \tau^{\nu} \partial_{\nu}h_{\rho \sigma} h^{\rho \sigma} - \partial_{\alpha}\tau_{\nu}\partial_{\rho}h_{\sigma \gamma} \tau^{\nu} \tau^{\rho} h^{\sigma \gamma}
- 2 \partial_{\alpha}\partial_{\nu} h_{\rho \sigma} \tau^{\rho} h^{\nu \sigma} - 2 \partial_{\mu}\partial_{\nu} h_{\alpha \sigma} \tau^{\mu} h^{\nu \sigma}
\nn \\ &&
+ 2 \partial_{\mu}\partial_{\nu} h_{\alpha \sigma} \tau^{\sigma} h^{\nu \mu} + 2 \partial_{\alpha}\partial_{\nu} h_{\rho \sigma} \tau^{\nu} h^{\rho \sigma}
- \frac32 \partial_{\alpha}h_{\nu \rho} \partial_{\sigma}h_{\gamma \epsilon} \tau^{\sigma} h^{\nu \gamma} h^{\rho \epsilon} + \frac12 \partial_{\alpha} h_{\nu \rho} \partial_{\sigma}h_{\gamma \epsilon} \tau^{\sigma} h^{\nu \rho} h^{\gamma \epsilon}
\nn \\ &&
- 2 \partial_{\alpha}\tau^{\nu}\partial_{\nu}\tau^{\rho} \tau_{\rho} - \partial_{\mu}\tau^{\nu} \partial_{\alpha}h_{\nu \sigma} h^{\mu \sigma} - 8 \partial_{\alpha}\varphi \partial_{\nu}\varphi \tau^{\nu}]
+ 2 \partial_{\beta}(\partial_{\rho} h_{\alpha \sigma} \tau^{\beta} h^{\rho \sigma} \Omega_{c} e^{-2\varphi})
\nn \\ &&
+ 4 \partial_{\beta}(\partial_{\mu} \tau^{\mu} \tau_{\alpha} \tau^{\beta} h^{\rho \sigma} \Omega_{c} e^{-2\varphi})
- 9 \partial_{\beta}(\partial_{\rho} \tau^{\sigma} \tau_{\alpha} \tau_{\sigma} \tau^{\beta} \tau^{\rho} \Omega_{c} e^{-2\varphi})
- 2 \partial_{\beta}(\partial_{\mu} \tau_{\nu} \tau_{\alpha} \tau^{\mu} \tau^{\beta} \tau^{\nu} \Omega_{c} e^{-2\varphi})
\nn \\ &&
+ 2 \partial_{\beta}(\partial_{\rho} h_{\gamma \sigma} \tau_{\alpha} \tau^{\beta} \tau^{\rho} h^{\gamma \sigma} \Omega_{c} e^{-2\varphi})
- \partial_{\beta}(\partial_{\alpha} h_{\rho \sigma} \tau^{\beta} h^{\rho \sigma} \Omega_{c} e^{-2\varphi})
+ 3 \partial_{\beta}(\partial_{\rho} \tau^{\sigma} h^{\beta \rho} h_{\alpha \sigma} \Omega_{c} e^{-2\varphi})
\nn \\ &&
- \partial_{\beta}(\partial_{\alpha}\tau^{\beta} \Omega_{c} e^{-2\varphi})
+ 2 \partial_{\beta}(\partial_{\alpha} \tau^{\rho} \tau_{\rho} \tau^{\beta} \Omega_{c} e^{-2\varphi})
+ 2 \partial_{\beta}(\partial_{\mu} \tau^{\beta} \tau_{\alpha} \tau^{\mu} \Omega_{c} e^{-2\varphi})
\nn \\ &&
+ \partial_{\beta}(\partial_{\rho}h_{\alpha \sigma} \tau^{\rho}h^{\beta \sigma} \Omega_{c} e^{-2\varphi})
+ 4 \partial_{\alpha}(\partial_{\nu} \tau^{\rho} \tau_{\rho} \tau^{\nu} \Omega_{c} e^{-2\varphi})
- 2 \partial_{\alpha}(\partial_{\nu} \tau^{\nu} \Omega_{c} e^{-2\varphi})
\nn \\ &&
+ 2 \partial_{\alpha}(\partial_{\mu} \tau_{\nu} \tau^{\mu} \tau^{\nu} \Omega_{c} e^{-2\varphi})
- \partial_{\alpha}(\partial_{\nu} h_{\rho \sigma} \tau^{\nu} h^{\rho \sigma} \Omega_{c} e^{-2\varphi}) = 0 \, .
\eea
The variations of the degenerate spatial metric and its inverse are respectively
\bea
\frac{\delta S}{\delta h_{\rho \sigma}} & = & \frac12 h^{\rho \sigma} \Omega_{c} e^{-2\varphi} L - \frac32 \Omega_{c} e^{-2\varphi} \partial_{\mu} \tau^{\rho} \partial_{\alpha}\tau^{\sigma} h^{\mu \alpha}
\nn \\ &&
+ 2 \partial_{\alpha}(\partial_{\mu}\tau^{\rho} \tau^{\mu} h^{\alpha \sigma} \Omega_{c} e^{-2\varphi})
- \partial_{\nu}(\partial_{\mu}\tau^{\mu} \tau^{\nu} h^{\rho \sigma} \Omega_{c} e^{-2\varphi})
\nn \\ &&
+ 2 \partial_{\alpha}(\partial_{\mu}\tau^{\nu} \tau_{\nu} \tau^{\mu} \tau^{\alpha} h^{\sigma \rho} \Omega_{c} e^{-2\varphi})
- \partial_{\nu}(\partial_{\mu}\tau^{\nu} \tau^{\mu} h^{\rho \sigma} \Omega_{c} e^{-2\varphi})
\nn \\ &&
+ \partial_{\alpha}(\partial_{\mu}\tau_{\nu} \tau^{\nu} \tau^{\mu} \tau^{\alpha} h^{\sigma \rho} \Omega_{c} e^{-2\varphi})
+ \frac32 \partial_{\mu}(\partial_{\alpha}h_{\gamma \epsilon} \tau^{\mu} \tau^{\alpha} h^{\sigma \gamma} h^{\rho \epsilon} \Omega_{c} e^{-2\varphi})
\nn \\ &&
- \frac12 \partial_{\mu}(\partial_{\alpha}h_{\gamma \epsilon} \tau^{\mu} \tau^{\alpha} h^{\sigma \rho} h^{\gamma \epsilon} \Omega_{c} e^{-2\varphi})
+ \partial_{\alpha}(\partial_{\mu} \tau^{\rho} \tau^{\alpha} h^{\mu \sigma} \Omega_{c} e^{-2\varphi})
\nn \\ &&
- 2 \partial_{\mu} \partial_{\nu}(\tau^{\mu} \tau^{\rho} h^{\nu \sigma} \Omega_{c} e^{-2\varphi})
+ \partial_{\mu} \partial_{\nu}(\tau^{\rho} \tau^{\sigma} h^{\nu \mu} \Omega_{c} e^{-2\varphi})
\nn \\ &&
+ \partial_{\mu} \partial_{\nu}(\tau^{\mu} \tau^{\nu} h^{\rho \sigma} \Omega_{c} e^{-2\varphi}) = 0 \, ,
\eea
and
\bea
\frac{\delta S}{\delta h^{\rho \sigma}} & = & \Omega_{c} e^{-2\varphi}[- 2 \partial_{\mu}\tau^{\nu}\partial_{\rho}h_{\nu \sigma} \tau^{\mu} + \partial_{\mu}\tau^{\mu}\partial_{\nu}h_{\rho \sigma} \tau^{\nu}
- 2 \partial_{\mu} \tau^{\nu} \partial_{\gamma} h_{\sigma \rho} \tau_{\nu} \tau^{\mu} \tau^{\gamma} 
\nn \\ &&
+ \partial_{\mu} \tau^{\nu} \partial_{\nu}h_{\rho \sigma} \tau^{\mu}
- \partial_{\mu} \tau_{\nu} \partial_{\gamma}h_{\sigma \rho} \tau^{\mu} \tau^{\nu} \tau^{\gamma} - 2 \partial_{\mu} \partial_{\rho}h_{\nu \sigma} \tau^{\mu} \tau^{\nu} 
\nn \\ &&
+ \partial_{\sigma} \partial_{\rho}h_{\nu \mu} \tau^{\mu} \tau^{\nu} + \partial_{\mu} \partial_{\nu} h_{\rho \sigma} \tau^{\mu} \tau^{\nu} 
\nn \\ &&
- \frac32 \partial_{\sigma}\tau^{\nu} \partial_{\rho} \tau^{\mu} h_{\mu \nu} - \frac32 \partial_{\mu} h_{\rho \nu} \partial_{\epsilon} h_{\gamma \sigma} \tau^{\mu} \tau^{\epsilon} h^{\nu \gamma} 
\nn \\ &&
+ \frac12 \partial_{\mu}h_{\rho \sigma} \partial_{\nu}h_{\gamma \epsilon} \tau^{\mu} \tau^{\nu} h^{\gamma \epsilon} - \partial_{\rho} \tau^{\nu} \partial_{\mu}h_{\nu \sigma} \tau^{\mu}] = 0 \, .
\eea
While the previous variations are generic, Lagrange multipliers should be incorporated in order to satisfy
\bea
\delta \tau_{\mu} h^{\mu \nu} + \tau_{\mu} \delta h^{\mu \nu} & = & 0 \, , \\
\delta \tau^{\mu} h_{\mu \nu} + \tau^{\mu} \delta h_{\mu \nu} & = & 0 \, , \\
\delta \tau_{\mu} \tau^{\mu} + \tau_{\mu} \delta \tau^{\mu} & = & 0 \, , \\
\delta \tau_{\mu} \tau^{\rho} + \tau_{\mu} \delta \tau^{\rho} + \delta h_{\mu \nu} h^{\nu \rho} + h_{\mu \nu} \delta h^{\nu \rho} & = & 0 \, .
\eea
Imposing the Carrollian ansatz on the relativistic equations of motion of the dilaton gravity theory we find:

\paragraph{Metric equations}
\bea
\Delta \hat g_{\mu \nu}^{(-4)} & = & - \frac12 \partial_{\rho}{\tau_{\sigma}} \partial_{\gamma}{\tau_{\epsilon}} \tau_{\mu} \tau_{\nu} h^{\rho \epsilon} h^{\sigma \gamma} + \frac12 \partial_{\rho}{\tau_{\sigma}} \partial_{\gamma}{\tau_{\epsilon}} \tau_{\mu} \tau_{\nu} h^{\rho \gamma} h^{\sigma \epsilon} = 0 \, ,
\eea
\bea
\Delta \hat g_{\mu \nu}^{(-2)} & = & \frac12 \partial_{\nu}{\tau_{\rho}} \partial_{\sigma}{h_{\gamma \epsilon}} \tau_{\mu}  h^{\rho \gamma} h^{\sigma \epsilon} + \frac12 \partial_{\nu}{\tau_{\rho}} \partial_{\sigma}{\tau_{\gamma}} \tau_{\mu} \tau^{\sigma}  h^{\rho \gamma} 
+ \frac12 \partial_{\nu}{\tau_{\rho}} \partial_{\sigma}{\tau_{\gamma}} \tau_{\mu} \tau^{\rho}  h^{\sigma \gamma} 
\nn \\ && 
+ \frac12 \partial_{\mu}{\tau_{\rho}} \partial_{\sigma}{h_{\gamma \epsilon}} \tau_{\nu}  h^{\rho \gamma} h^{\sigma \epsilon} 
+ \frac12 \partial_{\mu}{\tau_{\rho}} \partial_{\sigma}{\tau_{\gamma}} \tau_{\nu} \tau^{\sigma}  h^{\rho \gamma} + \frac12 \partial_{\mu}{\tau_{\rho}} \partial_{\sigma}{\tau_{\gamma}} \tau_{\nu} \tau^{\rho}  h^{\sigma \gamma} 
\nn \\ && 
- \frac12 \partial_{\rho}{\tau_{\nu}} \partial_{\sigma}{h_{\gamma \epsilon}} \tau_{\mu}  h^{\rho \gamma} h^{\sigma \epsilon} - \frac12 \partial_{\rho}{\tau_{\mu}} \partial_{\sigma}{h_{\gamma \epsilon}} \tau_{\nu}  h^{\rho \gamma} h^{\sigma \epsilon}  
- \frac12 \partial_{\rho}{\tau_{\nu}} \partial_{\sigma}{\tau_{\gamma}} \tau_{\mu} \tau^{\rho}  h^{\sigma \gamma} 
\nn \\ &&
- \frac12 \partial_{\rho}{\tau_{\mu}} \partial_{\sigma}{\tau_{\gamma}} \tau_{\nu} \tau^{\rho}  h^{\sigma \gamma} 
- \frac12 \partial_{\mu}{\tau_{\rho}} \partial_{\sigma}{\tau_{\nu}}  h^{\rho \sigma} - \frac12 \partial_{\mu \rho}{\tau_{\sigma}} \tau_{\nu}  h^{\rho \sigma} 
\nn \\ && 
+ \frac12 \partial_{\rho \sigma}{\tau_{\nu}} \tau_{\mu}  h^{\rho \sigma} + \frac12 \partial_{\rho}{\tau_{\mu}} \partial_{\sigma}{\tau_{\nu}}  h^{\rho \sigma}  
+ \frac12 \partial_{\rho \sigma}{\tau_{\mu}} \tau_{\nu}  h^{\rho \sigma} 
\nn \\ &&
+ \frac12 \partial_{\mu}{\tau_{\rho}} \partial_{\nu}{\tau_{\sigma}}  h^{\rho \sigma}  
- \frac12 \partial_{\nu}{\tau_{\rho}} \partial_{\sigma}{\tau_{\mu}}  h^{\rho \sigma} - \frac12 \partial_{\nu \rho}{\tau_{\sigma}} \tau_{\mu}  h^{\rho \sigma} 
\nn \\ && 
- \frac14 \partial_{\nu}{\tau_{\rho}} \partial_{\sigma}{h_{\gamma \epsilon}} \tau_{\mu}  h^{\rho \sigma} h^{\gamma \epsilon} - \frac12 \partial_{\nu}{\tau_{\rho}} \partial_{\sigma}{\tau_{\gamma}} \tau_{\mu} \tau^{\gamma}  h^{\rho \sigma} 
- \frac14 \partial_{\mu}{\tau_{\rho}} \partial_{\sigma}{h_{\gamma \epsilon}} \tau_{\nu}  h^{\rho \sigma} h^{\gamma \epsilon} 
\nn \\ &&
- \frac12 \partial_{\mu}{\tau_{\rho}} \partial_{\sigma}{\tau_{\gamma}} \tau_{\nu} \tau^{\gamma}  h^{\rho \sigma}  
+ \frac14 \partial_{\rho}{\tau_{\nu}} \partial_{\sigma}{h_{\gamma \epsilon}} \tau_{\mu}  h^{\rho \sigma} h^{\gamma \epsilon} + \frac14 \partial_{\rho}{\tau_{\mu}} \partial_{\sigma}{h_{\gamma \epsilon}} \tau_{\nu}  h^{\rho \sigma} h^{\gamma \epsilon} 
\nn \\ && 
+ \frac12 \partial_{\rho}{\tau_{\sigma}} \partial_{\gamma}{h_{\nu \epsilon}} \tau_{\mu}  h^{\rho \epsilon} h^{\sigma \gamma} + \frac12 \partial_{\rho}{\tau_{\sigma}} \partial_{\gamma}{h_{\mu \epsilon}} \tau_{\nu}  h^{\rho \epsilon} h^{\sigma \gamma}  
+ \partial_{\rho}{\tau_{\sigma}} \partial_{\gamma}{\tau_{\epsilon}} \tau_{\mu} \tau_{\nu} \tau^{\rho} \tau^{\epsilon}  h^{\sigma \gamma} 
\nn \\ &&
- \frac12 \partial_{\rho}{\tau_{\sigma}} \partial_{\gamma}{h_{\nu \epsilon}} \tau_{\mu}  h^{\rho \gamma} h^{\sigma \epsilon}  
- \frac12 \partial_{\rho}{\tau_{\sigma}} \partial_{\gamma}{h_{\mu \epsilon}} \tau_{\nu}  h^{\rho \gamma} h^{\sigma \epsilon} - \frac12 \partial_{\rho}{\tau_{\sigma}} \partial_{\gamma}{\tau_{\epsilon}} \tau_{\mu} \tau_{\nu} \tau^{\sigma} \tau^{\epsilon}  h^{\rho \gamma} 
\nn \\ && 
- \frac12 \partial_{\rho}{\tau_{\sigma}} \partial_{\gamma}{\tau_{\epsilon}} \tau_{\mu} \tau_{\nu} \tau^{\rho} \tau^{\gamma}  h^{\sigma \epsilon} + \partial_{\mu}{\tau_{\rho}} \partial_{\sigma}{\varphi} \tau_{\nu}  h^{\rho \sigma} 
+ \partial_{\nu}{\tau_{\rho}} \partial_{\sigma}{\varphi} \tau_{\mu}  h^{\rho \sigma} 
\nn \\ && 
- \partial_{\rho}{\tau_{\mu}} \partial_{\sigma}{\varphi} \tau_{\nu}  h^{\rho \sigma} 
- \partial_{\rho}{\tau_{\nu}} \partial_{\sigma}{\varphi} \tau_{\mu} h^{\rho \sigma} = 0 \, ,
\eea
\bea
\Delta \hat g_{\mu \nu}^{(0)} & = & - \frac12 \partial_{\nu}{h_{\mu \rho}} \partial_{\sigma}{h_{\gamma \epsilon}} h^{\rho \gamma} h^{\sigma \epsilon} - \frac12 \partial_{\rho}{\tau_{\sigma}} \partial_{\nu}{h_{\mu \gamma}} \tau^{\rho} h^{\sigma \gamma} + \frac12 \partial_{\rho}{\tau_{\sigma}} \partial_{\nu}{\tau^{\gamma}} h_{\mu \gamma} h^{\rho \sigma} 
\nn \\ &&
+ \frac12 \partial_{\nu}{\tau_{\rho}} \partial_{\sigma}{\tau^{\gamma}} \tau_{\mu} \tau^{\sigma} h_{\gamma \epsilon} h^{\rho \epsilon} + \frac12 \partial_{\nu}{\tau_{\mu}} \partial_{\rho}{\tau^{\sigma}} h_{\sigma \gamma} h^{\rho \gamma} + \frac12 \partial_{\nu}{\tau_{\rho}} \partial_{\sigma}{\tau^{\gamma}} \tau_{\mu} \tau^{\rho} h_{\gamma \epsilon} h^{\sigma \epsilon} 
\nn \\ &&
- \frac12 \partial_{\nu}{\tau_{\rho}} \partial_{\sigma}{\tau_{\gamma}} \tau_{\mu} \tau^{\rho} \tau^{\sigma} \tau^{\gamma} - \frac12 \partial_{\mu}{h_{\nu \rho}} \partial_{\sigma}{h_{\gamma \epsilon}} h^{\rho \gamma} h^{\sigma \epsilon} - \frac12 \partial_{\rho}{\tau_{\sigma}} \partial_{\mu}{h_{\nu \gamma}} \tau^{\rho} h^{\sigma \gamma} 
\nn \\ &&
+ \frac12 \partial_{\rho}{\tau_{\sigma}} \partial_{\mu}{\tau^{\gamma}} h_{\nu \gamma} h^{\rho \sigma} + \frac12 \partial_{\mu}{\tau_{\rho}} \partial_{\sigma}{\tau^{\gamma}} \tau_{\nu} \tau^{\sigma} h_{\gamma \epsilon} h^{\rho \epsilon} + \frac12 \partial_{\mu}{\tau_{\nu}} \partial_{\rho}{\tau^{\sigma}} h_{\sigma \gamma} h^{\rho \gamma} 
\nn \\ &&
+ \frac12 \partial_{\mu}{\tau_{\rho}} \partial_{\sigma}{\tau^{\gamma}} \tau_{\nu} \tau^{\rho} h_{\gamma \epsilon} h^{\sigma \epsilon} - \frac12 \partial_{\mu}{\tau_{\rho}} \partial_{\sigma}{\tau_{\gamma}} \tau_{\nu} \tau^{\rho} \tau^{\sigma} \tau^{\gamma} + \frac12 \partial_{\rho}{h_{\mu \nu}} \partial_{\sigma}{h_{\gamma \epsilon}} h^{\rho \gamma} h^{\sigma \epsilon} 
\nn \\ &&
+ \frac12 \partial_{\rho}{\tau_{\sigma}} \partial_{\gamma}{h_{\mu \nu}} \tau^{\rho} h^{\sigma \gamma} + \frac12 \partial_{\rho}{\tau_{\sigma}} \partial_{\gamma}{h_{\mu \nu}} \tau^{\gamma} h^{\rho \sigma} - \frac12 \partial_{\rho}{\tau_{\nu}} \partial_{\sigma}{\tau^{\gamma}} \tau_{\mu} \tau^{\sigma} h_{\gamma \epsilon} h^{\rho \epsilon} 
\nn \\ &&
- \frac12 \partial_{\rho}{\tau_{\mu}} \partial_{\sigma}{\tau^{\gamma}} \tau_{\nu} \tau^{\sigma} h_{\gamma \epsilon} h^{\rho \epsilon} - \frac12 \partial_{\rho}{\tau_{\nu}} \partial_{\sigma}{\tau^{\gamma}} \tau_{\mu} \tau^{\rho} h_{\gamma \epsilon} h^{\sigma \epsilon} - \frac12 \partial_{\rho}{\tau_{\mu}} \partial_{\sigma}{\tau^{\gamma}} \tau_{\nu} \tau^{\rho} h_{\gamma \epsilon} h^{\sigma \epsilon} 
\nn \\ &&
+ \frac12 \partial_{\rho}{\tau_{\nu}} \partial_{\sigma}{\tau_{\gamma}} \tau_{\mu} \tau^{\rho} \tau^{\sigma} \tau^{\gamma} + \frac12 \partial_{\rho}{\tau_{\mu}} \partial_{\sigma}{\tau_{\gamma}} \tau_{\nu} \tau^{\rho} \tau^{\sigma} \tau^{\gamma} + \frac12 \partial_{\mu \rho}{h_{\nu \sigma}} h^{\rho \sigma} 
\nn \\ &&
+ \frac12 \partial_{\mu \rho}{\tau_{\nu}} \tau^{\rho} + \frac12 \partial_{\mu}{\tau_{\rho}} \partial_{\sigma}{\tau_{\nu}} \tau^{\rho} \tau^{\sigma} + \frac12 \partial_{\mu \rho}{\tau_{\sigma}} \tau_{\nu} \tau^{\rho} \tau^{\sigma} 
\nn \\ &&
- \frac12 \partial_{\rho \sigma}{h_{\mu \nu}} h^{\rho \sigma} - \frac12 \partial_{\rho \sigma}{\tau_{\nu}} \tau_{\mu} \tau^{\rho} \tau^{\sigma} - \partial_{\rho}{\tau_{\mu}} \partial_{\sigma}{\tau_{\nu}} \tau^{\rho} \tau^{\sigma} 
\nn \\ &&
- \frac12 \partial_{\rho \sigma}{\tau_{\mu}} \tau_{\nu} \tau^{\rho} \tau^{\sigma} + \frac14 \partial_{\mu}{h_{\rho \sigma}} \partial_{\nu}{h_{\gamma \epsilon}} h^{\rho \gamma} h^{\sigma \epsilon} - \frac12 \partial_{\mu}{\tau_{\rho}} \partial_{\nu}{\tau^{\sigma}} h_{\sigma \gamma} h^{\rho \gamma} 
\nn \\ &&
- \frac12 \partial_{\nu}{\tau_{\rho}} \partial_{\mu}{\tau^{\sigma}} h_{\sigma \gamma} h^{\rho \gamma} - \frac12 \partial_{\mu \nu}{h_{\rho \sigma}} h^{\rho \sigma} - \partial_{\mu \nu}{\tau_{\rho}} \tau^{\rho} 
\nn \\ &&
+ \frac12 \partial_{\nu \rho}{h_{\mu \sigma}} h^{\rho \sigma} + \frac12 \partial_{\nu \rho}{\tau_{\mu}} \tau^{\rho} + \frac12 \partial_{\nu}{\tau_{\rho}} \partial_{\sigma}{\tau_{\mu}} \tau^{\rho} \tau^{\sigma} 
\nn \\ &&
+ \frac12 \partial_{\nu \rho}{\tau_{\sigma}} \tau_{\mu} \tau^{\rho} \tau^{\sigma} + \frac14 \partial_{\nu}{h_{\mu \rho}} \partial_{\sigma}{h_{\gamma \epsilon}} h^{\rho \sigma} h^{\gamma \epsilon} + \frac12 \partial_{\rho}{\tau_{\sigma}} \partial_{\nu}{h_{\mu \gamma}} \tau^{\sigma} h^{\rho \gamma} 
\nn \\ &&
+ \frac14 \partial_{\nu}{\tau_{\mu}} \partial_{\rho}{h_{\sigma \gamma}} \tau^{\rho} h^{\sigma \gamma} + \frac14 \partial_{\nu}{\tau_{\rho}} \partial_{\sigma}{h_{\gamma \epsilon}} \tau_{\mu} \tau^{\rho} \tau^{\sigma} h^{\gamma \epsilon} + \frac14 \partial_{\mu}{h_{\nu \rho}} \partial_{\sigma}{h_{\gamma \epsilon}} h^{\rho \sigma} h^{\gamma \epsilon} 
\nn \\ &&
+ \frac12 \partial_{\rho}{\tau_{\sigma}} \partial_{\mu}{h_{\nu \gamma}} \tau^{\sigma} h^{\rho \gamma} + \frac14 \partial_{\mu}{\tau_{\nu}} \partial_{\rho}{h_{\sigma \gamma}} \tau^{\rho} h^{\sigma \gamma} + \frac14 \partial_{\mu}{\tau_{\rho}} \partial_{\sigma}{h_{\gamma \epsilon}} \tau_{\nu} \tau^{\rho} \tau^{\sigma} h^{\gamma \epsilon} 
\nn \\ &&
- \frac14 \partial_{\rho}{h_{\mu \nu}} \partial_{\sigma}{h_{\gamma \epsilon}} h^{\rho \sigma} h^{\gamma \epsilon} - \frac12 \partial_{\rho}{\tau_{\sigma}} \partial_{\gamma}{h_{\mu \nu}} \tau^{\sigma} h^{\rho \gamma} - \frac14 \partial_{\rho}{\tau_{\nu}} \partial_{\sigma}{h_{\gamma \epsilon}} \tau_{\mu} \tau^{\rho} \tau^{\sigma} h^{\gamma \epsilon} 
\nn \\ &&
- \frac14 \partial_{\rho}{\tau_{\mu}} \partial_{\sigma}{h_{\gamma \epsilon}} \tau_{\nu} \tau^{\rho} \tau^{\sigma} h^{\gamma \epsilon} - \frac12 \partial_{\rho}{h_{\mu \sigma}} \partial_{\gamma}{h_{\nu \epsilon}} h^{\rho \epsilon} h^{\sigma \gamma} + \frac12 \partial_{\rho}{\tau_{\sigma}} \partial_{\gamma}{\tau^{\epsilon}} \tau_{\mu} \tau^{\rho} h_{\nu \epsilon} h^{\sigma \gamma} 
\nn \\ &&
- \frac12 \partial_{\rho}{\tau_{\mu}} \partial_{\sigma}{h_{\nu \gamma}} \tau^{\sigma} h^{\rho \gamma} - \frac12 \partial_{\rho}{\tau_{\sigma}} \partial_{\gamma}{h_{\nu \epsilon}} \tau_{\mu} \tau^{\sigma} \tau^{\gamma} h^{\rho \epsilon} - \frac12 \partial_{\rho}{\tau_{\nu}} \partial_{\sigma}{h_{\mu \gamma}} \tau^{\sigma} h^{\rho \gamma} 
\nn \\ &&
- \frac12 \partial_{\rho}{\tau_{\sigma}} \partial_{\gamma}{h_{\mu \epsilon}} \tau_{\nu} \tau^{\sigma} \tau^{\gamma} h^{\rho \epsilon} + \frac12 \partial_{\rho}{\tau_{\sigma}} \partial_{\gamma}{\tau^{\epsilon}} \tau_{\nu} \tau^{\rho} h_{\mu \epsilon} h^{\sigma \gamma} + \frac12 \partial_{\rho}{h_{\mu \sigma}} \partial_{\gamma}{h_{\nu \epsilon}} h^{\rho \gamma} h^{\sigma \epsilon} 
\nn \\ &&
- \frac12 \partial_{\rho}{\tau_{\mu}} \partial_{\sigma}{\tau^{\gamma}} h_{\nu \gamma} h^{\rho \sigma} - \frac12 \partial_{\rho}{\tau_{\sigma}} \partial_{\gamma}{\tau^{\epsilon}} \tau_{\mu} \tau^{\sigma} h_{\nu \epsilon} h^{\rho \gamma} + \frac12 \partial_{\rho}{\tau_{\sigma}} \partial_{\gamma}{h_{\nu \epsilon}} \tau_{\mu} \tau^{\rho} \tau^{\gamma} h^{\sigma \epsilon} 
\nn \\ &&
- \frac12 \partial_{\rho}{\tau_{\nu}} \partial_{\sigma}{\tau^{\gamma}} h_{\mu \gamma} h^{\rho \sigma} - \frac12 \partial_{\rho}{\tau_{\sigma}} \partial_{\gamma}{\tau^{\epsilon}} \tau_{\nu} \tau^{\sigma} h_{\mu \epsilon} h^{\rho \gamma} + \frac12 \partial_{\rho}{\tau_{\sigma}} \partial_{\gamma}{h_{\mu \epsilon}} \tau_{\nu} \tau^{\rho} \tau^{\gamma} h^{\sigma \epsilon} 
\nn \\ &&
+ 2 \partial_{\mu \nu}{\varphi} - \partial_{\rho}{\varphi} \partial_{\mu}{h_{\nu \sigma}} h^{\rho \sigma} - \partial_{\mu}{\tau_{\nu}} \partial_{\rho}{\varphi} \tau^{\rho} 
- \partial_{\mu}{\tau_{\rho}} \partial_{\sigma}{\varphi} \tau_{\nu} \tau^{\rho} \tau^{\sigma} - \partial_{\rho}{\varphi} \partial_{\nu}{h_{\mu \sigma}} h^{\rho \sigma} - \partial_{\nu}{\tau_{\mu}} \partial_{\rho}{\varphi} \tau^{\rho} 
\nn \\ &&
- \partial_{\nu}{\tau_{\rho}} \partial_{\sigma}{\varphi} \tau_{\mu} \tau^{\rho} \tau^{\sigma} + \partial_{\rho}{\varphi} \partial_{\sigma}{h_{\mu \nu}} h^{\rho \sigma} + \partial_{\rho}{\tau_{\mu}} \partial_{\sigma}{\varphi} \tau_{\nu} \tau^{\rho} \tau^{\sigma} + \partial_{\rho}{\tau_{\nu}} \partial_{\sigma}{\varphi} \tau_{\mu} \tau^{\rho} \tau^{\sigma} = 0 \, ,
\eea

\bea
\Delta \hat g^{(2)}_{\mu \nu} & = & - \frac12 \partial_{\rho}{\tau^{\sigma}} \partial_{\nu}{h_{\mu \gamma}} \tau^{\rho}  h_{\sigma \epsilon} h^{\gamma \epsilon} + \frac12 \partial_{\nu}{\tau^{\rho}} \partial_{\sigma}{\tau^{\gamma}}  h_{\mu \rho} h_{\gamma \epsilon} h^{\sigma \epsilon} - \frac12 \partial_{\rho}{\tau_{\sigma}} \partial_{\nu}{\tau^{\gamma}} \tau^{\rho} \tau^{\sigma}  h_{\mu \gamma} 
\nn \\ &&
- \frac12 \partial_{\rho}{\tau^{\sigma}} \partial_{\mu}{h_{\nu \gamma}} \tau^{\rho}  h_{\sigma \epsilon} h^{\gamma \epsilon} + \frac12 \partial_{\mu}{\tau^{\rho}} \partial_{\sigma}{\tau^{\gamma}}  h_{\nu \rho} h_{\gamma \epsilon} h^{\sigma \epsilon} - \frac12 \partial_{\rho}{\tau_{\sigma}} \partial_{\mu}{\tau^{\gamma}} \tau^{\rho} \tau^{\sigma}  h_{\nu \gamma} 
\nn \\ &&
+ \frac12 \partial_{\rho}{\tau^{\sigma}} \partial_{\gamma}{h_{\mu \nu}} \tau^{\rho}  h_{\sigma \epsilon} h^{\gamma \epsilon} + \frac12 \partial_{\rho}{\tau^{\sigma}} \partial_{\gamma}{h_{\mu \nu}} \tau^{\gamma}  h_{\sigma \epsilon} h^{\rho \epsilon} - \frac12 \partial_{\rho}{\tau_{\sigma}} \partial_{\gamma}{h_{\mu \nu}} \tau^{\rho} \tau^{\sigma} \tau^{\gamma}  
\nn \\ &&
- \frac12 \partial_{\mu \rho}{h_{\nu \sigma}} \tau^{\rho} \tau^{\sigma}  + \frac12 \partial_{\rho \sigma}{h_{\mu \nu}} \tau^{\rho} \tau^{\sigma}  - \frac12 \partial_{\mu}{\tau^{\rho}} \partial_{\nu}{\tau^{\sigma}}  h_{\rho \gamma} h_{\sigma \epsilon} h^{\gamma \epsilon} 
\nn \\ &&
+ \frac12 \partial_{\mu \nu}{h_{\rho \sigma}} \tau^{\rho} \tau^{\sigma}  - \frac12 \partial_{\nu \rho}{h_{\mu \sigma}} \tau^{\rho} \tau^{\sigma}  + \frac14 \partial_{\nu}{\tau^{\rho}} \partial_{\sigma}{h_{\gamma \epsilon}} \tau^{\sigma}  h_{\mu \rho} h^{\gamma \epsilon} 
\nn \\ &&
+ \frac14 \partial_{\mu}{\tau^{\rho}} \partial_{\sigma}{h_{\gamma \epsilon}} \tau^{\sigma}  h_{\nu \rho} h^{\gamma \epsilon} + \frac14 \partial_{\rho}{h_{\mu \nu}} \partial_{\sigma}{h_{\gamma \epsilon}} \tau^{\rho} \tau^{\sigma}  h^{\gamma \epsilon} - \frac12 \partial_{\rho}{\tau^{\sigma}} \partial_{\gamma}{h_{\mu \epsilon}} \tau^{\gamma}  h_{\nu \sigma} h^{\rho \epsilon} 
\nn \\ &&
- \frac12 \partial_{\rho}{\tau^{\sigma}} \partial_{\gamma}{h_{\nu \epsilon}} \tau^{\gamma}  h_{\mu \sigma} h^{\rho \epsilon} - \frac12 \partial_{\rho}{\tau^{\sigma}} \partial_{\gamma}{\tau^{\epsilon}}  h_{\mu \sigma} h_{\nu \epsilon} h^{\rho \gamma} - \frac12 \partial_{\rho}{h_{\mu \sigma}} \partial_{\gamma}{h_{\nu \epsilon}} \tau^{\rho} \tau^{\gamma}  h^{\sigma \epsilon} 
\nn \\ &&
- \partial_{\mu}{\tau^{\rho}} \partial_{\sigma}{\varphi} \tau^{\sigma}  h_{\nu \rho} - \partial_{\nu}{\tau^{\rho}} \partial_{\sigma}{\varphi} \tau^{\sigma}  h_{\mu \rho} - \partial_{\rho}{\varphi} \partial_{\sigma}{h_{\mu \nu}} \tau^{\rho} \tau^{\sigma} = 0 \, . 
\eea

\paragraph{Dilaton equations}
\bea
\Delta \hat{\phi}^{(-2)} & = & - \frac12 \partial_{\mu}{\tau_{\nu}} \partial_{\rho}{\tau_{\sigma}} h^{\mu \sigma} h^{\nu \rho} + \frac12 \partial_{\mu}{\tau_{\nu}} \partial_{\rho}{\tau_{\sigma}} h^{\mu \rho} h^{\nu \sigma} \, ,
\eea
\bea
\Delta \hat{\phi}^{(0)} & = & - \partial_{\mu}{h_{\nu \rho}} \partial_{\sigma}{h_{\gamma \epsilon}} h^{\mu \nu} h^{\rho \gamma} h^{\sigma \epsilon} - 2 \partial_{\mu}{\tau_{\nu}} \partial_{\rho}{h_{\sigma \gamma}} \tau^{\mu} h^{\nu \sigma} h^{\rho \gamma} + 2 \partial_{\mu}{\tau_{\nu}} \partial_{\rho}{\tau^{\sigma}} h_{\sigma \gamma} h^{\mu \nu} h^{\rho \gamma}
\nn \\ &&
- 2 \partial_{\mu}{\tau_{\nu}} \partial_{\rho}{\tau_{\sigma}} \tau^{\mu} \tau^{\rho} h^{\nu \sigma} + \partial_{\mu}{h_{\nu \rho}} \partial_{\sigma}{h_{\gamma \epsilon}} h^{\mu \nu} h^{\rho \sigma} h^{\gamma \epsilon} + \partial_{\mu}{\tau_{\nu}} \partial_{\rho}{h_{\sigma \gamma}} \tau^{\mu} h^{\nu \rho} h^{\sigma \gamma} 
\nn \\ &&
+ \partial_{\mu}{\tau_{\nu}} \partial_{\rho}{h_{\sigma \gamma}} \tau^{\rho} h^{\mu \nu} h^{\sigma \gamma} + 2 \partial_{\mu}{\tau_{\nu}} \partial_{\rho}{h_{\sigma \gamma}} \tau^{\nu} h^{\mu \sigma} h^{\rho \gamma} + \partial_{\mu \nu}{h_{\rho \sigma}} h^{\mu \rho} h^{\nu \sigma} 
\nn \\ &&
+ 2 \partial_{\mu}{\tau_{\nu}} \partial_{\rho}{\tau_{\sigma}} \tau^{\mu} \tau^{\sigma} h^{\nu \rho} + 2 \partial_{\mu \nu}{\tau_{\rho}} \tau^{\mu} h^{\nu \rho} - \partial_{\mu \nu}{h_{\rho \sigma}} h^{\mu \nu} h^{\rho \sigma} 
\nn \\ &&
- 2 \partial_{\mu \nu}{\tau_{\rho}} \tau^{\rho} h^{\mu \nu} + \frac34 \partial_{\mu}{h_{\nu \rho}} \partial_{\sigma}{h_{\gamma \epsilon}} h^{\mu \sigma} h^{\nu \gamma} h^{\rho \epsilon} - 3 \partial_{\mu}{\tau_{\nu}} \partial_{\rho}{\tau^{\sigma}} h_{\sigma \gamma} h^{\mu \rho} h^{\nu \gamma} 
\nn \\ &&
- \frac14 \partial_{\mu}{h_{\nu \rho}} \partial_{\sigma}{h_{\gamma \epsilon}} h^{\mu \sigma} h^{\nu \rho} h^{\gamma \epsilon} - \partial_{\mu}{\tau_{\nu}} \partial_{\rho}{h_{\sigma \gamma}} \tau^{\nu} h^{\mu \rho} h^{\sigma \gamma} - \frac12 \partial_{\mu}{h_{\nu \rho}} \partial_{\sigma}{h_{\gamma \epsilon}} h^{\mu \gamma} h^{\nu \sigma} h^{\rho \epsilon} 
\nn \\ &&
+ \partial_{\mu}{\tau_{\nu}} \partial_{\rho}{\tau^{\sigma}} h_{\sigma \gamma} h^{\mu \gamma} h^{\nu \rho} - \partial_{\mu}{\tau_{\nu}} \partial_{\rho}{h_{\sigma \gamma}} \tau^{\rho} h^{\mu \sigma} h^{\nu \gamma} + 4 \partial_{\mu}{\varphi} \partial_{\nu}{\varphi} h^{\mu \nu} = 0 \, .
\eea
\bea
\Delta \hat{\phi}^{(2)} & = & - 2 \partial_{\mu}{\tau^{\nu}} \partial_{\rho}{h_{\sigma \gamma}} \tau^{\mu}  h_{\nu \epsilon} h^{\rho \sigma} h^{\gamma \epsilon} + \partial_{\mu}{\tau^{\nu}} \partial_{\rho}{\tau^{\sigma}}  h_{\nu \gamma} h_{\sigma \epsilon} h^{\mu \gamma} h^{\rho \epsilon} - 2 \partial_{\mu}{\tau_{\nu}} \partial_{\rho}{\tau^{\sigma}} \tau^{\mu} \tau^{\nu}  h_{\sigma \gamma} h^{\rho \gamma} 
\nn \\ &&
+ \partial_{\mu}{\tau^{\nu}} \partial_{\rho}{h_{\sigma \gamma}} \tau^{\mu}  h_{\nu \epsilon} h^{\rho \epsilon} h^{\sigma \gamma} + \partial_{\mu}{\tau^{\nu}} \partial_{\rho}{h_{\sigma \gamma}} \tau^{\rho}  h_{\nu \epsilon} h^{\mu \epsilon} h^{\sigma \gamma} - \partial_{\mu}{\tau_{\nu}} \partial_{\rho}{h_{\sigma \gamma}} \tau^{\mu} \tau^{\nu} \tau^{\rho}  h^{\sigma \gamma} 
\nn \\ &&
- 2 \partial_{\mu \nu}{h_{\rho \sigma}} \tau^{\mu} \tau^{\rho}  h^{\nu \sigma} + \partial_{\mu \nu}{h_{\rho \sigma}} \tau^{\rho} \tau^{\sigma}  h^{\mu \nu} + \partial_{\mu \nu}{h_{\rho \sigma}} \tau^{\mu} \tau^{\nu}  h^{\rho \sigma} 
\nn \\ &&
- \frac34 \partial_{\mu}{h_{\nu \rho}} \partial_{\sigma}{h_{\gamma \epsilon}} \tau^{\mu} \tau^{\sigma}  h^{\nu \gamma} h^{\rho \epsilon} - \frac32 \partial_{\mu}{\tau^{\nu}} \partial_{\rho}{\tau^{\sigma}}  h_{\nu \gamma} h_{\sigma \epsilon} h^{\mu \rho} h^{\gamma \epsilon} + \frac14 \partial_{\mu}{h_{\nu \rho}} \partial_{\sigma}{h_{\gamma \epsilon}} \tau^{\mu} \tau^{\sigma}  h^{\nu \rho} h^{\gamma \epsilon} 
\nn \\ &&
+ \frac12 \partial_{\mu}{\tau^{\nu}} \partial_{\rho}{\tau^{\sigma}}  h_{\nu \gamma} h_{\sigma \epsilon} h^{\mu \epsilon} h^{\rho \gamma} - \partial_{\mu}{\tau^{\nu}} \partial_{\rho}{h_{\sigma \gamma}} \tau^{\rho}  h_{\nu \epsilon} h^{\mu \sigma} h^{\gamma \epsilon} - 4 \partial_{\mu}{\varphi} \partial_{\nu}{\varphi} \tau^{\mu} \tau^{\nu} = 0 \, .
\eea


\begin{thebibliography}{}
\bibitem{Poincare1}
N. Sen Gupta, “On an Analogue of the Galileo Group,”
Nuovo Cim. 54 (1966) 512,  DOI:
10.1007/BF02740871 .

\bibitem{Poincare2}
J. Levy-Leblond, “Une nouvelle limite non-relativiste
du group de Poincare,” Ann.Inst.Henri Poincare 3
(1965) 1.

\bibitem{BH1}
R. F. Penna, “Near-horizon Carroll symmetry and black
hole Love numbers,” arXiv:1812.05643 [hep-th].
\bibitem{BH2}
L. Donnay and C. Marteau, “Carrollian Physics at the
Black Hole Horizon,” Class. Quant. Grav. 36 no. 16,
(2019) 165002, arXiv:1903.09654 [hep-th].
\bibitem{BH3}
D. Hansen, N. A. Obers, G. Oling, and B. T. Søgaard,
“Carroll Expansion of General Relativity,” SciPost
Phys. 13 no. 3, (2022) 055, arXiv:2112.12684
[hep-th].
\bibitem{BH4}
A. Pérez, “Asymptotic symmetries in Carrollian
theories of gravity,” JHEP 12 (2021) 173,
arXiv:2110.15834 [hep-th].
\bibitem{BH5} 
J. Redondo-Yuste and L. Lehner, “Non-linear black hole
dynamics and Carrollian fluids,” JHEP 02 (2023) 240,
arXiv:2212.06175 [gr-qc].
[8] F. Ecker, D. Grumiller, J. Hartong, A. Pérez,
S. Prohazka, and R. Troncoso, “Carroll black holes,”
arXiv:2308.10947 [hep-th].

\bibitem{NearHorizon1}
A. Bagchi, A. Banerjee, J. Hartong, E. Have, K. S. Kolekar, "Strings near black holes are Carrollian", Phys.Rev.D 110 (2024) 8, 086009.

\bibitem{NearHorizon2}
A. Bagchi, A. Banerjee, J. Hartong, E. Have, K. S. Kolekar, "Strings near black holes are Carrollian. Part II", JHEP 11 (2024) 024.

\bibitem{NearHorizon3}
S. Hüsnügil and Luis Lehner, "Sourced Carrollian fluids dual to black hole horizons", Phys.Rev.D 112 (2025) 10, 104043

\bibitem{FlatHolo1}
A. Bagchi, “Correspondence between Asymptotically
Flat Spacetimes and Nonrelativistic Conformal Field
Theories,” Phys.Rev.Lett. 105 (2010) 171601.
\bibitem{FlatHolo2}
G. Barnich, A. Gomberoff, and H. A. Gonzalez, “The
Flat limit of three dimensional asymptotically anti-de
Sitter spacetimes,” Phys. Rev. D 86 (2012) 024020,
arXiv:1204.3288 [gr-qc].
\bibitem{FlatHolo3}
A. Bagchi, S. Detournay, R. Fareghbal, and J. Simon,
“Holography of 3D Flat Cosmological Horizons,”
Phys.Rev.Lett. 110 no. 14, (2013) 141302,
arXiv:1208.4372 [hep-th].
\bibitem{FlatHolo4}
A. Bagchi and R. Fareghbal, “BMS/GCA Redux:
Towards Flatspace Holography from Non-Relativistic
Symmetries,” JHEP 1210 (2012) 092, arXiv:1203.5795
[hep-th].
\bibitem{FlatHolo5}
J. Hartong, “Holographic Reconstruction of 3D Flat
Space-Time,” JHEP 10 (2016) 104, arXiv:1511.01387
[hep-th].
\bibitem{FlatHolo6}
A. Saha, “Carrollian approach to 1 + 3D flat
holography,” JHEP 06 (2023) 051, arXiv:2304.02696
[hep-th].
\bibitem{FlatHolo7}
A. Bagchi, P. Dhivakar, and S. Dutta, “Holography in
Flat Spacetimes: the case for Carroll,”
arXiv:2311.11246 [hep-th].

\bibitem{FlatHolo8}
A. Fontanella and O. Payne, “A Carroll Limit of AdS/CFT: A Triality with Flat Space
Holography?,” arXiv:2508.10085 [hep-th]

\bibitem{FlatHolo9}
G. Arenas-Henriquez, L. Ciambelli, F. Diaz, W. Jia, D. Rivera-Betancour, "Radiation in fluid/gravity and the flat limit", JHEP 01 (2026) 086.


\bibitem{FlatHolo10}
G. Poulias and S. Vandoren, "On Carroll partition functions and flat space holography", JHEP 06 (2025) 232.

\bibitem{FlatHolo11}
F. Diaz, "Nonperfect Carrollian Fluids Through Holography", e-Print: 2602.00396 [hep-th].

\bibitem{FlatHolo12}
S. Fredenhagen, S. Prohazka, R. Tiefenbacher, "Carrollian quantum states and flat space holography", e-Print: 2604.22745 [hep-th].

\bibitem{CeleHolo1}
J. Figueroa-O’Farrill, E. Have, S. Prohazka, and
J. Salzer, “Carrollian and celestial spaces at infinity,”
JHEP 09 (2022) 007, arXiv:2112.03319 [hep-th].

\bibitem{CeleHolo2}
L. Donnay, A. Fiorucci, Y. Herfray, and R. Ruzziconi,
“Carrollian Perspective on Celestial Holography,” Phys.
Rev. Lett. 129 no. 7, (2022) 071602, arXiv:2202.04702
[hep-th].
\bibitem{CeleHolo3}
A. Bagchi, S. Banerjee, R. Basu, and S. Dutta,
“Scattering Amplitudes: Celestial and Carrollian,”
Phys. Rev. Lett. 128 no. 24, (2022) 241601,
arXiv:2202.08438 [hep-th].

\bibitem{CeleHolo4}
L. Donnay, A. Fiorucci, Y. Herfray, and R. Ruzziconi,
“Bridging Carrollian and celestial holography,” Phys.
Rev. D 107 no. 12, (2023) 126027, arXiv:2212.12553
[hep-th].
\bibitem{CeleHolo5}
L. Mason, R. Ruzziconi, and A. Yelleshpur Srikant,
“Carrollian Amplitudes and Celestial Symmetries,”
arXiv:2312.10138 [hep-th].

\bibitem{Null1}
B. Chen, R. Liu, H. Sun, and Y.-f. Zheng, “Constructing Carrollian field theories from null
reduction,” JHEP 11 (2023) 170, arXiv:2301.06011 [hep-th].
\bibitem{Null2}
A. Sharma, “Studies on Carrollian quantum field theories,” Class. Quant. Grav. 43 no. 4,
(2026) 045006, arXiv:2502.00487 [hep-th].
\bibitem{Null3} 
S. Majumdar, “Carroll theories from Lorentzian light-cone theories,” JHEP 02 (2026) 258,
arXiv:2507.03081 [hep-th].
\bibitem{Null4}
S. Majumdar, A. Sharma, and S. Singha, “Carroll fermions from null reduction: A case of
good and bad fermions,” arXiv:2605.05334 [hep-th].

\bibitem{Eff1}
L. Ciambelli, C. Marteau, A. C. Petkou, P. M.
Petropoulos, and K. Siampos, “Flat holography and
Carrollian fluids,” JHEP 07 (2018) 165,
arXiv:1802.06809 [hep-th].

\bibitem{Eff2}
A. Campoleoni, L. Ciambelli, C. Marteau, P. M.
Petropoulos, and K. Siampos, “Two-dimensional fluids
and their holographic duals,” Nucl. Phys. B 946 (2019)
114692, arXiv:1812.04019 [hep-th].
\bibitem{Eff3}
A. C. Petkou, P. M. Petropoulos, D. R. Betancour, and
K. Siampos, “Relativistic fluids, hydrodynamic frames
and their Galilean versus Carrollian avatars,” JHEP 09
(2022) 162, arXiv:2205.09142 [hep-th].
\bibitem{Eff4}
L. Freidel and P. Jai-akson, “Carrollian hydrodynamics
from symmetries,” Class. Quant. Grav. 40 no. 5, (2023)
055009, arXiv:2209.03328 [hep-th].
\bibitem{Eff5}
J. de Boer, J. Hartong, N. A. Obers, W. Sybesma, and
S. Vandoren, “Carroll stories,” arXiv:2307.06827
[hep-th].
\bibitem{Eff6}
J. Armas and E. Have, “Carrollian fluids and
spontaneous breaking of boost symmetry,”
arXiv:2308.10594 [hep-th].

\bibitem{Eff7}
E. A. Bergshoeff, P. Concha, O. Fierro, E. Rodríguez, J. Rosseel, "A conformal approach to Carroll gravity", JHEP 07 (2025) 075.

\bibitem{Eff8}
A. Argandoña, A. Guijosa, S. Patiño-López, "De Sitter Holography and Carrollian brane theories", JHEP 10 (2025) 133, e-Print: 2507.06147 [hep-th]

\bibitem{Eff9}
    L. Ciambelli and P. Jai-akson, "Foundations of Carrollian geometry", Phys.Rept. 1188 (2026) 1-51.
    
\bibitem{Eff10}
    H. Afshar and M. Ahmadi-Jahmani, "Scaling Symmetry and Carrollian Gravity", e-Print: 2512.20736 [hep-th].
    
\bibitem{Eff11}
    H.T. Özer and Aytül Filiz, "Holonomies and Boundary Symmetries in the Discrete BF Formulation of Carroll Dilaton Gravity", e-Print: 2606.25499 [hep-th].

\bibitem{NR1}
U. H. Danielsson, A. Guijosa, and M. Kruczenski, “IIA/B, wound and
wrapped,” JHEP 10 (2000) 020, arXiv:hep-th/0009182.

\bibitem{NR2}
J. Gomis and H. Ooguri, “Nonrelativistic closed string theory,” J. Math. Phys., vol. 42,
pp. 3127–3151, 2001.

\bibitem{NR3}
U. H. Danielsson, A. Guijosa, and M. Kruczenski, “Newtonian gravitons and
D-brane collective coordinates in wound string theory,” JHEP 03 (2001) 041, arXiv:hep-th/0012183.

\bibitem{NRST1}
E. A. Bergshoeff, K. T. Grosvenor, C. Simsek, and Z. Yan, “An Action for Extended String
Newton-Cartan Gravity,” JHEP, vol. 01, p. 178, 2019.

\bibitem{NRST2}
T. Harmark, J. Hartong, L. Menculini, N. A. Obers, and G. Oling, “Relating non-relativistic
string theories,” JHEP, vol. 11, p. 071, 2019.

\bibitem{NRST3}
E. A. Bergshoeff, J. Gomis, J. Rosseel, C. Simsek, and Z. Yan, “String Theory and String Newton-Cartan Geometry,” J. Phys. A, vol. 53, no. 1, p. 014001, 2020.

\bibitem{NRST4}
Z. Yan and M. Yu, “Background Field Method for Nonlinear Sigma Models in Nonrelativistic
String Theory,” JHEP, vol. 03, p. 181, 2020.

\bibitem{NRST5}
B. Julia and H. Nicolai, “Null-killing vector dimensional reduction and galilean geometrodynamics,” Nuclear Physics B, vol. 439, no. 1, pp. 291 – 323, 1995.

\bibitem{NRST6}
E. Bergshoeff, J. Gomis, and Z. Yan, “Nonrelativistic String Theory and T-Duality,” JHEP,
vol. 11, p. 133, 2018.

\bibitem{NRST7}
J. Kluson, “Remark About Non-Relativistic String in Newton-Cartan Background and Null
Reduction,” JHEP, vol. 05, p. 041, 2018.

\bibitem{NRST8}
J. Gomis, J. Oh, and Z. Yan, “Nonrelativistic String Theory in Background
Fields,” JHEP 10 (2019) 101, arXiv:hep-th/1905.07315.

\bibitem{NRST9}
A. D. Gallegos, U. Gursoy, and N. Zinnato, “Torsional Newton Cartan gravity
from non-relativistic strings,” JHEP 09 (2020) 172, arXiv:hep-th/1906.01607.

\bibitem{NRST10} L. Bidussi, T. Harmark, J. Hartong, N. A. Obers, and G. Oling, Torsional string
Newton-Cartan geometry for non-relativistic strings, JHEP {\bf 02} (2022) 116,
arXiv:hep-th/2107.00642.

\bibitem{New} 
E. A. Bergshoeff, L. Romano, J. Rosseel, E. Simón-Félix, S. Zeko, "From Galilei to Euclidean Carroll and the Alice Particle: The Times They Are a-Changin'", e-Print: 2607.05115 [hep-th].

\bibitem{Sugra1}
F. Ali, L. Ravera, "N-extended Chern-Simons Carrollian supergravities in 2+1 spacetime dimensions", JHEP 02 (2020) 128.

\bibitem{Sugra2}
L. Ravera, U. Zorba, "Carrollian and non-relativistic Jackiw–Teitelboim supergravity", Eur.Phys.J.C 83 (2023) 2, 107.

\bibitem{Sugra3}
    D. Grumiller, L. Montecchio, M. Shams Nejati, "Carroll dilaton supergravity in two dimensions", JHEP 12 (2024) 005

\bibitem{Sugra4}
B. Chen and Z. Hu, "Carrollian superstring in the flipped vacuum", Phys.Rev.D 112 (2025) 4, 046005.

\bibitem{Sugra5}
I. Bulunur, O. Ergec, O. Kasikci, M. Ozkan, M. Salih Zog, "A Twisted Origin for Magnetic Carroll Supersymmetry", e-Print: 2603.28269 [hep-th].

\bibitem{Sugra6}
M. Henneaux, "Carroll supergravities", e-Print: 2607.08329 [hep-th].

\bibitem{GS}
  M.~B.~Green and J.~H.~Schwarz,
  ``Anomaly Cancellation in Supersymmetric D=10 Gauge Theory and Superstring Theory,''
  Phys.\ Lett.\ B {\bf 149} (1984) 117.


\bibitem{EL}
E. Lescano, "Curvatures and Non-metricities in the Non-Relativistic Limit of Bosonic Supergravity", e-Print: 2601.03342 [hep-th].


\bibitem{Eheterotic}
E. Lescano, ``A Non-Relativistic Limit for Heterotic Supergravity and its Gauge Lagrangian'', Nucl.Phys.B 1028 2026, 117487, 2502.08711 [hep-th].

\bibitem{Boost}
E. Bergshoeff, J. Gomis, B. Rollier, J. Rosseel, T. ter Veldhuis, "Carroll versus Galilei Gravity", JHEP 03 (2017) 165.

\bibitem{NSNS}
E.~A. Bergshoeff, J.~Lahnsteiner, L.~Romano, J.~Rosseel and C.Simsek,
  ´´A non-relativistic limit of NS-NS gravity´´, JHEP {\bf 06}
  (2021) 021, arXiv:hep-th/2102.06974.

  \bibitem{BandR}
E. A. Bergshoeff and L. Romano, Non-relativistic heterotic string theory, JHEP 01
(2024) 146, arXiv:hep-th/2310.19716.

\bibitem{LO}
E. Lescano and D. Osten, Non-relativistic limits of bosonic and heterotic Double Field Theory, JHEP 07 (2024) 286, arXiv:hep-th/2405.10362.

\bibitem{V2citation1}
M.M. Sheikh-Jabbari, H. Yavartanoo, "On the Consistency of Null Strings Literature: The Tale of an Overlooked Symmetry", 2605.12414 [hep-th].

\bibitem{V2citation2}
M.M. Sheikh-Jabbari, H. Yavartanoo, "Null Strings Gauged and Reloaded, I: Null Strings Have Carroll-Weyl Gauge Symmetry", 2605.25817 [hep-th].

\bibitem{V2citation3}
M.M. Sheikh-Jabbari, H. Yavartanoo, "Null Strings Gauged and Reloaded, II: Consistent Classical Treatment of the Null Strings", 2605.26822 [hep-th]

\bibitem{V2citation4}
U. Lindstrom, "Symmetries of tensionless strings", 2605.26185 [hep-th].

\bibitem{V2citation5}
S. Duary, S. Maji, "Path integral quantization of null bosonic strings with Carroll-Weyl ghosts", 2606.04999 [hep-th].

\bibitem{V2citation6}
U. Lindstrom, "The conformal null string in d+2 and 
d dimensions", 2606.22498 [hep-th].

\bibitem{V2citation7}
Ida M. Rasulian, M.M. Sheikh-Jabbari, H. Yavartanoo, "Null-strings Gauged, Reloaded and Quantized, I: Canonical Quantization in the Light-Cone Gauge", 2607.02970 [hep-th].

 \bibitem{NRDFT1}
S. M. Ko, C. Melby-Thompson, R. Meyer, and J.-H. Park, ``Dynamics of
Perturbations in Double Field Theory \& Non-Relativistic String Theory", JHEP {\bf 12}
(2015) 144, [arXiv:1508.01121].

\bibitem{NRDFT2}
K. Morand and J.-H. Park, ``Classification of non-Riemannian doubled-yet-gauged
spacetime", Eur. Phys. J. C {\bf 77} (2017), no. 10 685, [arXiv:1707.03713]. [Erratum:
Eur.Phys.J.C 78, 901 (2018)].

\bibitem{NRDFT3}
K. Cho and J.-H. Park, ``Remarks on the non-Riemannian sector in Double Field
Theory", Eur. Phys. J. C {\bf 80} (2020), no. 2 101, [arXiv:1909.10711].

\bibitem{NRDFT4}
A. D. Gallegos, U. Gursoy, S. Verma, and N. Zinnato, ``Non-Riemannian gravity
actions from double field theory", JHEP {\bf 06} (2021) 173, [arXiv:2012.07765].

\bibitem{NRDFT5}
C. D. A. Blair, G. Oling, and J.-H. Park, ``Non-Riemannian isometries from double
field theory", JHEP {\bf 04} (2021) 072, [arXiv:2012.07766].

\bibitem{DFT1}
  C.~Hull and B.~Zwiebach, ``Double Field Theory,'' JHEP {\bf 0909} (2009) 099 [arXiv:0904.4664 [hep-th]].

\bibitem{DFT2}
 C.~Hull and B.~Zwiebach, ``The Gauge algebra of double field theory and Courant brackets,'' JHEP {\bf 0909} (2009) 090 [arXiv:0908.1792 [hep-th]].

\bibitem{DFT3}
  O.~Hohm, C.~Hull and B.~Zwiebach, ``Background independent action for double field theory,'' JHEP {\bf 1007} (2010) 016 [arXiv:1003.5027 [hep-th]].

\bibitem{DFT4}
  O.~Hohm, C.~Hull and B.~Zwiebach, ``Generalized metric formulation of double field theory,''  JHEP {\bf 1008} (2010) 008  [arXiv:1006.4823 [hep-th]].


 \bibitem{ReviewDFT1}
  G.~Aldazabal, D.~Marques and C.~Nunez,
  ``Double Field Theory: A Pedagogical Review,''
  Class.\ Quant.\ Grav.\  {\bf 30}, 163001 (2013)
  [arXiv:1305.1907 [hep-th]].

\bibitem{ReviewDFT2}
   O.~Hohm, D.~Lüst and B.~Zwiebach,
  ``The Spacetime of Double Field Theory: Review, Remarks, and Outlook,''
  Fortsch.\ Phys.\  {\bf 61}, 926 (2013)
  [arXiv:1309.2977 [hep-th]].

\bibitem{ReviewDFT3}
J-H.Park, ``Gravitational Core of Double Field Theory: Lecture Notes'', e-Print: 2505.10163.

\bibitem{HeteroticAP1}
E. Lescano, "Trivialization of the gravitational Green-Schwarz transformation in the nonrelativistic limit of string theory", Phys.Rev.D 113 (2026) 4, L041902.

\bibitem{HeteroticAP2}
E. Lescano, "Gravitational four-derivative corrections in non-relativistic heterotic supergravity and the SO(8) Green-Schwarz mechanism", e-Print: 2508.09250 [hep-th].

\bibitem{FigueroaNew}
    J. Figueroa-O'Farrill, E. Have, N. A. Obers, e-Print: 2509.04397 [hep-th]

\bibitem{BLP2}
R. Ballesteros, E. Lescano and S. Patiño-López, work in progress.
\end{thebibliography}
\end{document}